\documentclass[aps,prb,twocolumn,superscriptaddress,amsmath,amssymb,floatfix]{revtex4-2}

\usepackage{graphicx}
\usepackage{bm}
\usepackage{dcolumn}
\usepackage{hyperref}
\usepackage{xcolor}

\newcommand{\rr}{\mathbf r}
\newcommand{\RR}{\mathbf R}
\newcommand{\kk}{\mathbf k}
\newcommand{\qq}{\mathbf q}

\newcommand{\ml}{\mathrm{ML}}
\newcommand{\DFT}{\mathrm{DFT}}

\newcommand{\loc}{\mathrm{loc}}

\newcommand{\tpsi}{\widetilde{\psi}}
\newcommand{\tH}{\widetilde{H}}
\newcommand{\tS}{\widetilde{S}}
\newcommand{\tv}{\widetilde{v}}
\newcommand{\tp}{\widetilde{p}}
\newcommand{\tphi}{\widetilde{\phi}}
\newcommand{\FIXME}[1]{\textcolor{red}{\ifmmode\mathbf{[#1]}\else\textbf{[#1]}\fi}}

\begin{document}

\title{Machine Learning Local Potentials for Accelerated Electron-Phonon Interactions Calculations within the Projector Augmented-Wave Framework}

\author{Yi Xia}
\email{yimaverickxia@gmail.com}
\email{yxia@pdx.edu}
\affiliation{Department of Mechanical and Materials Engineering, Portland State University, Portland, Oregon 97201, USA}

\date{\today}

\begin{abstract}
Electron–phonon interactions govern carrier dynamics, transport, and many optical and quantum phenomena in solids, but finite-displacement calculations within the projector augmented-wave (PAW) framework require up to \(6N\) self-consistent supercell calculations for an \(N\)-atom system, making them costly for large, low-symmetry, and disordered materials. We introduce MLLocP, a machine learning strategy that learns the self-consistent local potential on real-space grids and supplies its displacement derivatives to the evaluation of all-electron PAW electron–phonon matrix elements. Analysis of the PAW decomposition identifies the local-potential contribution as a natural target for machine learning, while the remaining PAW quantities are retained within the established VASP/PHELEL framework. Using targeted grid-point sampling and feature-diversity selection for efficient model training, we validate MLLocP for elemental Cu, polar BAs, and chemically disordered Cu\(_3\)Au. The learned potentials and their resulting displacement derivatives closely reproduce direct density functional theory calculations, yielding transport coefficients within approximately 3.5\% for BAs mobility, 8\% for Cu conductivity, and 11\% for Cu\(_3\)Au conductivity. For the 32-atom Cu\(_3\)Au special quasirandom structure, using 40 sampled configurations reduces the number of self-consistent calculations by about fivefold relative to the 192 displaced structures required directly; using 20 or 10 configurations increases the acceleration to approximately 10- and 20-fold, with conductivity errors of 14.3\% and 18.6\%, respectively. MLLocP thus provides a scalable route to PAW electron–phonon calculations in complex materials while preserving the underlying all-electron formalism.
\end{abstract}

\maketitle

\section{Introduction}

Electron-phonon interactions provide one of the central microscopic mechanisms by which electrons exchange energy and momentum with the vibrating lattice~\cite{Grimvall1981,Allen1982,Giustino2017}. They control the temperature dependence of electronic bands~\cite{giustino2010electron}, carrier mobility~\cite{ponce2020first}, superconducting pairing~\cite{bardeen1957theory}, hot-carrier relaxation~\cite{bernardi2014ab}, and nonradiative recombination~\cite{alkauskas2014first}, and they also contribute to the decoherence of quantum states in emerging quantum materials and devices~\cite{chirolli2008decoherence}. Predictive calculations of electron-phonon coupling are therefore essential for connecting atomic-scale structure to transport, optical, and thermal properties across metals, semiconductors, insulators, and complex functional materials.

First-principles electron-phonon calculations have advanced substantially over the past two decades. Density functional perturbation theory (DFPT)~\cite{Baroni2001} provides an efficient route to the linear variation of the self-consistent Kohn-Sham potential~\cite{kohn1965self,Hhhenberg1964inhomogeneous} with respect to collective phonon perturbations. Combined with maximally localized Wannier functions~\cite{marzari1997maximally,Marzari2012}, it enables interpolation of electron-phonon matrix elements from coarse to dense Brillouin-zone meshes~\cite{Giustino2007,Ponce2016}, which has become a standard approach for computing carrier scattering and dynamics~\cite{Zhou2016,cao2018dominant,coulter2019uncovering,xia2019high,brunin2020electron,jhalani2020piezoelectric,Zhou2021,Protik2022,wang2026accurate}, superconductivity~\cite{margine2013anisotropic}, and temperature-dependent band structures~\cite{Giustino2017, lee2023electron}. A complementary route is the supercell finite-displacement approach~\cite{kaasbjerg2012phonon,Gunst2016}, in which atomic displacements are imposed explicitly and the derivative of the electronic Hamiltonian is constructed from differences between self-consistent calculations. This real-space strategy is conceptually straightforward, naturally interfaces with supercell descriptions of defects and disorder, and has recently been developed in the projector augmented-wave (PAW) formalism~\cite{Blochl1994}, including the PHELEL finite-displacement approach~\cite{Chaput2019,Engel2020,Engel2022,PHELEL}, as implemented in the Vienna \textit{Ab initio} Simulation Package (VASP) ecosystem~\cite{Blochl1994,KresseHafner1993,Kresse1996,KresseJoubert1999}.

Despite this progress, the computational cost of electron-phonon calculations remains a severe bottleneck for large, low-symmetry, and chemically disordered systems. In a conventional central-difference supercell calculation, each inequivalent atomic Cartesian displacement requires two additional self-consistent electronic-structure calculations~\cite{PHELEL}. For an $N$-atom cell without symmetry, the number of displaced configurations therefore scales as $6N$, before the additional costs of dense electronic state sampling, band summations, and transport calculations are included. This scaling makes direct first-principles electron-phonon calculations difficult for alloys, high-entropy materials, defect-rich semiconductors, interfaces, and other structurally complex systems in which electron-phonon physics may be essential.

ML offers a promising route to reduce this cost by replacing repeated electronic structure calculations with data-driven surrogate models. Existing approaches have achieved remarkable success in learning interatomic potentials~\cite{Unke2021,chen2022universal,batatia2022mace}, electronic Hamiltonians in localized bases~\cite{DeepH2022,HamGNN2023}, charge densities~\cite{Jorgensen2022, Feng2025}, wave functions~\cite{Schutt2019}, and density responses~\cite{Jorgensen2022,Feng2025}. For electron-phonon problems, one natural strategy is to learn a localized electronic Hamiltonian as a function of atomic structure and then differentiate the Hamiltonian with respect to atomic displacements, as pursued in frameworks such as DeepH~\cite{DeepH2022, li2024first} and HamGNN~\cite{HamGNN2023, zhong2024accelerating}. Such Hamiltonian learning approaches are powerful, but their accuracy depends on the quality, transferability, and completeness of the chosen localized basis. This dependence can be challenging in transition metal systems, compounds with semicore states, disordered metallic alloys, or materials requiring a high-fidelity PAW treatment of valence and near-core electronic structure.

The PAW method provides an alternative and widely used first-principles framework~\cite{Blochl1994}. It combines the computational efficiency of a smooth pseudo wave function with an all-electron reconstruction inside atom-centered augmentation spheres~\cite{Blochl1994,KresseJoubert1999}. Within the PAW formalism, the electron-phonon matrix element can be expressed in terms of smooth pseudo wave functions, the self-consistent local potential, PAW strength matrices, projector functions, and one-center reconstruction terms~\cite{Chaput2019,Engel2020,Engel2022}. This structure suggests a different ML strategy: instead of learning a localized Hamiltonian in a chosen atomic orbital basis, one may learn the self-consistent local potential field itself. The local potential is a scalar real-space field in the absence of spin-orbit coupling and can be treated similarly to the electron density. Recent ML models for electron density have shown that real-space fields can be learned efficiently by combining local atomic descriptors with targeted grid-point sampling and feature-selection strategies~\cite{Jorgensen2022,Feng2025}.

In this work, we develop the formal and computational basis for machine learning local potentials (MLLocP) to accelerate electron-phonon calculations. We focus on three questions. First, which terms in the PAW electron-phonon matrix element depend on self-consistent electronic response, and which are determined primarily by the fixed PAW dataset and analytic translations of atom-centered functions? Second, which PAW quantities are most naturally represented by ML models? Third, how can a learned local potential field be integrated into a finite-displacement PAW electron-phonon workflow by taking advantage of existing implementations?

The main conclusion of our analysis is that the self-consistent local potential is the most direct first target for ML. It is high-dimensional, configuration-dependent, and expensive to obtain from repeated self-consistent calculations, yet it is represented on a smooth real-space grid and has the same locality structure exploited in modern charge-density learning. We therefore formulate a model that maps atomic positions and chemical species to a scalar PAW local potential on the real-space grid. Displacement derivatives of this learned potential are constructed with the PHELEL finite-displacement approach and supplied to the PAW electron-phonon expression, while the pseudo wave functions, PAW projectors, overlap matrices, and remaining augmentation quantities, including terms that involve derivatives of the projectors, are handled by VASP for the corresponding primitive cell.

The remainder of the paper is organized as follows. Section~\ref{sec:theory} reviews the PAW electron-phonon matrix element and analyzes the learnability of each contribution by ML. Section~\ref{sec:ml_problem} defines the ML problem for the local potential, including grid-point sampling algorithms and the evaluation of displacement derivatives within the PHELEL/VASP workflow. Section~\ref{sec:workflow} describes the proposed VASP-MLLocP-PHELEL workflow. Section~\ref{sec:computational_details} introduces the three validation systems: elemental Cu, polar BAs, and disordered Cu$_3$Au, together with the DFT and ML protocols. Results and discussion are presented in Sec.~\ref{sec:results}.

\section{Theory and methods}
\label{sec:theory}

\subsection{Electron-phonon matrix elements}

For a phonon mode with wave vector $\qq$ and branch $\nu$, the first-order electron-phonon matrix element between Bloch states $|\psi_{n\kk}\rangle$ and $|\psi_{m,\kk+\qq}\rangle$ can be written as~\cite{Giustino2007,Giustino2017}
\begin{equation}
  g_{mn\nu}(\kk,\qq)=
  \sum_{\kappa\alpha}
  \left(\frac{\hbar}{2M_\kappa \omega_{\nu\qq}}\right)^{1/2}
  e_{\kappa\alpha}^{\nu\qq}
  \left\langle \psi_{m,\kk+\qq}\left|
  \partial_{\kappa\alpha,\qq} \hat{H}
  \right|\psi_{n\kk}\right\rangle,
  \label{eq:g_general}
\end{equation}
where $\kappa$ labels atoms in the primitive cell, $\alpha$ is a Cartesian direction, $M_\kappa$ is the atomic mass, $\omega_{\nu\qq}$ and $e_{\kappa\alpha}^{\nu\qq}$ are phonon frequencies and eigenvectors, and $\partial_{\kappa\alpha,\qq}\hat{H}$ denotes the first-order derivative of the Kohn-Sham Hamiltonian associated with the collective phonon perturbation. In a real-space finite-displacement supercell approach, the building block is the derivative with respect to the displacement of atom $\kappa$ in Cartesian direction $\alpha$,
\begin{equation}
  g_{mn}^{\kappa\alpha}(\kk,\kk')=
  \left\langle \psi_{m\kk'}\left|
  \frac{\partial \hat{H}}{\partial R_{\kappa\alpha}}
  \right|\psi_{n\kk}\right\rangle,
  \label{eq:g_atomic}
\end{equation}
which is subsequently Fourier transformed and contracted with phonon eigenvectors.

Within the PAW method~\cite{Blochl1994, KresseJoubert1999}, the all-electron wave function is related to a smooth pseudo wave function by
\begin{equation}
  |\psi_{n\kk}\rangle = \hat{T} |\tpsi_{n\kk}\rangle,
  \qquad
  \hat{T}=1+\sum_i \left(|\phi_i\rangle-|\tphi_i\rangle\right)\langle \tp_i|,
  \label{eq:paw_transformation}
\end{equation}
where $|\phi_i\rangle$, $|\tphi_i\rangle$, and $|\tp_i\rangle$ are all-electron partial waves, pseudo partial waves, and PAW projector functions, respectively. The compound index $i=(\kappa_i,n_i,l_i,m_i)$ labels the atom and angular-momentum channel. The generalized PAW eigenvalue problem is
\begin{equation}
  \tH |\tpsi_{n\kk}\rangle = \epsilon_{n\kk} \tS |\tpsi_{n\kk}\rangle,
  \label{eq:paw_gen_eig}
\end{equation}
with
\begin{align}
  \tS &= 1 + \sum_{ij} |\tp_i\rangle Q_{ij}\langle \tp_j|, \label{eq:paw_overlap}\\
  \tH &= -\frac{\hbar^2}{2m_e}\nabla^2 + \tv(\rr)
  + \sum_{ij}|\tp_i\rangle D_{ij}\langle \tp_j| .
  \label{eq:paw_hamiltonian}
\end{align}
Here $Q_{ij}$ is the PAW overlap augmentation charges, $D_{ij}$ is the PAW strength parameter, and $\tv(\rr)$ is the smooth local potential on the real-space grid. In the following we write $V_\loc(\rr)\equiv\tv(\rr)$ for the smooth self-consistent local potential, and denote its ML and DFT values by $V_\loc^\ml$ and $V_\loc^\DFT$, respectively.

Following the PAW electron-phonon formulation, the atomic displacement derivative in Eq.~\eqref{eq:g_atomic} can be decomposed into four contributions~\cite{Chaput2019,Engel2020,Engel2022},
\begin{equation}
  g_{mn}^{\kappa\alpha} =
  g_{mn}^{(V),\kappa\alpha}
  +g_{mn}^{(D),\kappa\alpha}
  +g_{mn}^{(P),\kappa\alpha}
  +g_{mn}^{(R),\kappa\alpha}.
  \label{eq:paw_decomposition}
\end{equation}
The local potential term is
\begin{equation}
  g_{mn}^{(V),\kappa\alpha}=\left\langle\tpsi_m\left|
  \frac{\partial \tv}{\partial R_{\kappa\alpha}}
  \right|\tpsi_n\right\rangle,
  \label{eq:gV}
\end{equation}
where band and wave-vector indices are suppressed for compactness. The PAW strength term is
\begin{equation}
  g_{mn}^{(D),\kappa\alpha}=\sum_{ij}
  \langle \tpsi_m|\tp_i\rangle
  \frac{\partial D_{ij}}{\partial R_{\kappa\alpha}}
  \langle \tp_j|\tpsi_n\rangle.
  \label{eq:gD}
\end{equation}
The projector derivative term is
\begin{align}
  g_{mn}^{(P),\kappa\alpha}=&
  \sum_{ij}
  \left\langle \tpsi_m\left|\frac{\partial \tp_i}{\partial R_{\kappa\alpha}}\right.\right\rangle
  \left(D_{ij}-\epsilon_n Q_{ij}\right)
  \langle \tp_j|\tpsi_n\rangle \nonumber\\
  &+\sum_{ij}
  \langle \tpsi_m|\tp_i\rangle
  \left(D_{ij}-\epsilon_m Q_{ij}\right)
  \left\langle \left.\frac{\partial \tp_j}{\partial R_{\kappa\alpha}}\right|\tpsi_n\right\rangle,
  \label{eq:gP}
\end{align}
where $\epsilon_n$ and $\epsilon_m$ denote the corresponding Kohn-Sham eigenvalues. Finally, the reconstruction term is
\begin{widetext}
\begin{align}
  g_{mn}^{(R),\kappa\alpha}=&
  -\left(\epsilon_m-\epsilon_n\right)
  \sum_{ij}\langle \tpsi_m|\tp_i\rangle
  \left[
  \left\langle \phi_i\left|\frac{\partial \phi_j}{\partial R_{\kappa\alpha}}\right.\right\rangle
  -
  \left\langle \tphi_i\left|\frac{\partial \tphi_j}{\partial R_{\kappa\alpha}}\right.\right\rangle
  \right]
  \langle \tp_j|\tpsi_n\rangle .
  \label{eq:gR}
\end{align}
\end{widetext}
Eqs.~\eqref{eq:gV}--\eqref{eq:gR} are the starting point for identifying which parts of the PAW electron-phonon calculation can be accelerated by ML.

\subsection{Which PAW quantities should be learned?}

The four terms in Eq.~\eqref{eq:paw_decomposition} differ substantially in physical content, locality, dimensionality, and dependence on self-consistent electronic response. This distinction is important because it determines whether an ML model should target a real-space field, a low-dimensional augmentation matrix, or an analytic atom-centered function.

The term $g^{(V)}$ contains the derivative of  $\tv(\rr)$ with respect to atomic displacements. This potential includes the smooth local ionic contribution and the self-consistent Hartree and exchange-correlation response. Therefore, $\partial \tv/\partial R_{\kappa\alpha}$ is not a fixed pseudopotential quantity; it must be obtained from self-consistent calculations in a finite-displacement approach. At the same time, $\tv(\rr)$ is represented on a regular real-space grid and is a scalar field in collinear calculations without spin-orbit coupling. These features make $g^{(V)}$ the most natural and highest impact target for ML.

The term $g^{(D)}$ contains the derivative of the PAW strength matrix. The matrices $D_{ij}$ are localized within PAW augmentation spheres and have a much smaller dimensionality than the real-space potential. They include one-center contributions and depend on the self-consistent potential and density through PAW augmentation quantities. Thus, their displacement derivatives can also require finite-displacement information. In practice, however, we find that $g^{(D)}$ is negligible compared with $g^{(V)}$ for the systems studied here (see detailed discussion in Sec.~\ref{sec:gD_vs_gV}), which is also consistent with earlier PAW finite-displacement analyses of electron-phonon matrix elements~\cite{Chaput2019,Engel2020,PHELEL}. We therefore exclude $g^{(D)}$ in our ML task.

The term $g^{(P)}$ arises from derivatives of the PAW projectors with respect to atomic positions. The projector functions themselves are fixed by the PAW dataset and translate rigidly with the atoms. Their derivatives are therefore analytic derivatives of localized atom-centered functions. $g^{(P)}$ also depends on $D_{ij}$, $Q_{ij}$, the eigenvalues, and the pseudo-wave-functions, all of which can be obtained from a single self-consistent calculation for the undisplaced primitive structure. Therefore, this term is not a natural target for ML because much of its structure is already encoded in the PAW dataset. 

The term $g^{(R)}$ contains derivatives of all-electron and pseudo partial waves and represents an all-electron reconstruction correction associated with the generalized PAW eigenvalue problem. As in $g^{(P)}$, its atom-centered functions are fixed by the PAW dataset and their displacement derivatives follow from translations of the augmentation spheres. 
Since this term is tied closely to PAW dataset and quantities from self-consistent calculation of the undisplaced primitive structure, it is better retained within the established PAW implementation rather than replaced by a separate surrogate model at this stage.

This analysis motivates the central approximation explored in this work:
\begin{equation}
  \frac{\partial \tv(\rr;\RR)}{\partial R_{\kappa\alpha}}
  \approx
  \frac{\partial \tv_{\ml}(\rr;\RR)}{\partial R_{\kappa\alpha}},
  \label{eq:central_approx}
\end{equation}
where $\RR=\{\RR_\kappa\}$ denotes the atomic configuration and $\tv_{\ml}$ is an ML local potential. The other sandwich quantities for the primitive cell are obtained directly from the DFT calculation, including the smooth PAW pseudo wave functions, eigenvalues, the PAW overlap augmentation charges $Q_{ij}$, and the PAW strength parameters $D_{ij}$. This separation allows the ML model to replace repeated self-consistent calculations for the local potential field while preserving the DFT-derived PAW quantities used to evaluate the final electron-phonon matrix elements.

\subsection{Short-range and long-range components of the local potential}

For nonpolar metals and short-range screened systems, the displacement derivative of the local potential is expected to decay rapidly in real space~\cite{kohn1996density,prodan2005nearsightedness}. For polar semiconductors and insulators, however, long-range dipole fields associated with Born effective charges and the electronic dielectric response produce nonanalytic electron-phonon couplings near the Brillouin-zone center~\cite{Verdi2015,Sjakste2015,Engel2022,brunin2020electron,jhalani2020piezoelectric}. In a general formulation for polar materials, a purely local ML model would therefore be trained on the short-range component of the potential after subtracting an analytic long-range contribution constructed from Born effective charges and the high-frequency dielectric tensor, with that long-range term added back when evaluating polar matrix elements~\cite{Verdi2015,Sjakste2015,Chaput2019}.

In the present work we do not apply this short-range/long-range separation. For metallic Cu and Cu$_3$Au the polar long-range correction is absent. For polar BAs, the 64-atom supercell is comparatively small, so that the relevant atomic environments lie within the cutoff radius of the local ML model; we therefore train directly on the full self-consistent local potential. As shown in Sec.~\ref{sec:results}, this direct treatment still yields accurate local potentials, derivatives, and electron-phonon observables for BAs, thanks to the interpolation scheme implemented in VASP~\cite{Chaput2019,Engel2022}, which separates the short- and long-range contributions to the derivatives of the local potential. Explicit short-range/long-range separation remains a natural extension for larger polar supercells where long-range fields extend beyond the ML cutoff.

\section{ML formulation}
\label{sec:ml_problem}

\subsection{Local potential learning problem}

For each atomic configuration $s$, a self-consistent DFT calculation provides a smooth local potential on a uniform real-space grid,
\begin{equation}
  \mathcal{D}=\left\{\RR^{(s)},\, Z^{(s)},\, \left[\rr_g^{(s)},\tv^{(s)}(\rr_g)\right]_{g=1}^{N_g}\right\}_{s=1}^{N_s},
  \label{eq:dataset}
\end{equation}
where $Z^{(s)}$ denotes chemical species and $N_g$ is the number of grid points. The goal is to learn a model
\begin{equation}
  \tv_{\ml}(\rr_g;\RR,Z)=F_\theta\left(\mathcal{E}(\rr_g;\RR,Z)\right),
  \label{eq:ml_mapping}
\end{equation}
where $\mathcal{E}(\rr_g;\RR,Z)$ is the local atomic environment around grid point $\rr_g$ and $F_\theta$ is a neural network with parameters $\theta$. We adapt the field-induced recursively embedded atom neural network (FIREANN) architecture~\cite{Zhang2019,Zhang2021,Zhang2022,ZhangJiang2023,Feng2025} from ML of the electron density on real-space grids to ML of the self-consistent local potential in the present study. Each grid point is treated as a virtual (ghost) atom whose environment is described by embedded atom density (EAD) features. The $n$th EAD feature of grid point $g$ is the squared norm of a symmetry-adapted linear combination of Gaussian-type orbitals centered on the neighboring real atoms within a cutoff radius $r_c$,
\begin{equation}
  \rho_{n}(\rr_g)=\sum_{l_x+l_y+l_z=L}\frac{L!}{l_x! l_y! l_z!}
  \left|\sum_{\kappa}^{N_c} c_\kappa\,\varphi^{n}_{l_x l_y l_z}(\rr_g-\RR_\kappa)\right|^2,
  \label{eq:ead_feature}
\end{equation}
where the sum over $\kappa$ runs over neighboring atoms, $L$ is the total angular momentum, $\varphi^{n}$ are contracted Gaussian-type orbitals modulated by a cosine cutoff function, and the coefficients $c_\kappa$ carry the chemical species dependence. Because $\mathcal{E}=\{\rho_n\}$ is built from relative vectors $\rr_g-\RR_\kappa$ and squared orbital sums, it is invariant under rotations, translations, and permutations of like atoms, so the scalar output $\tv_\ml$ inherits the correct symmetry of the local potential. Nonlocal and higher many-body effects are incorporated by making the coefficients $c_\kappa$ the output of an atomic neural network that is updated over $T$ message-passing iterations~\cite{Zhang2021}. Here, we map the network output to $V_\loc$ rather than to the charge density.

A dense three-dimensional grid contains many points with redundant local environments, so training on every grid point is inefficient and may bias the model toward slowly varying interstitial regions. We therefore adopt the efficient real-space grid-point sampling strategy developed within the FIREANN density framework~\cite{Feng2025}: targeted sampling first selects points with large field values or large field variations, and linearly independent feature-based sampling then retains only those candidates whose EAD feature vectors [Eq.~\eqref{eq:ead_feature}] are not linearly reproducible from already accepted points. This two-step procedure reduces the number of training grid points while preserving information-rich regions and diverse atomic environments.

\subsection{Displacement derivatives}

Once the local-potential model is trained, the atomic displacement derivative needed in Eq.~\eqref{eq:gV} is obtained with the PHELEL finite-displacement approach~\cite{Chaput2019,Engel2022,PHELEL}: the learned potentials $\tv_{\ml}$ on positively and negatively displaced configurations replace the corresponding self-consistent DFT local potentials, and PHELEL constructs
\begin{equation}
  \frac{\partial \tv_{\ml}}{\partial R_{\kappa\alpha}}
  \approx
  \frac{\tv_{\ml}(\RR+\delta\mathbf{e}_{\kappa\alpha})
  -\tv_{\ml}(\RR-\delta\mathbf{e}_{\kappa\alpha})}{2\delta}
  \label{eq:ml_fd}
\end{equation}
by the same central-difference procedure used in a conventional PAW finite-displacement calculation. The resulting derivative field is inserted into the PAW local-potential matrix element,
\begin{equation}
  g_{mn,\ml}^{(V),\kappa\alpha}
  =\left\langle\tpsi_m\left|
  \frac{\partial \tv_{\ml}}{\partial R_{\kappa\alpha}}
  \right|\tpsi_n\right\rangle,
  \label{eq:gV_ml}
\end{equation}
which is evaluated within the VASP/PHELEL workflow using the same plane-wave and real-space-grid conventions as the underlying PAW code.

The full approximate PAW matrix element can be written as
\begin{equation}
  g_{mn,\ml}^{\kappa\alpha}
  = g_{mn,\ml}^{(V),\kappa\alpha}
  + g_{mn}^{(D),\kappa\alpha}
  + g_{mn}^{(P),\kappa\alpha}
  + g_{mn}^{(R),\kappa\alpha},
  \label{eq:full_g_ml}
\end{equation}
where the remaining contributions, in particular the projector-derivative term $g^{(P)}$ and the reconstruction term $g^{(R)}$, are computed internally within VASP from the PAW projectors, partial waves, and pseudo wave functions. In practice $g^{(D)}$ is negligible for the systems studied here (Sec.~\ref{sec:gD_vs_gV}). This strategy is intentionally conservative: only the expensive self-consistent local-potential field is replaced, while the established PAW machinery is preserved for the final all-electron matrix elements.

\begin{figure}[t]
    \includegraphics[width = 0.75\linewidth]{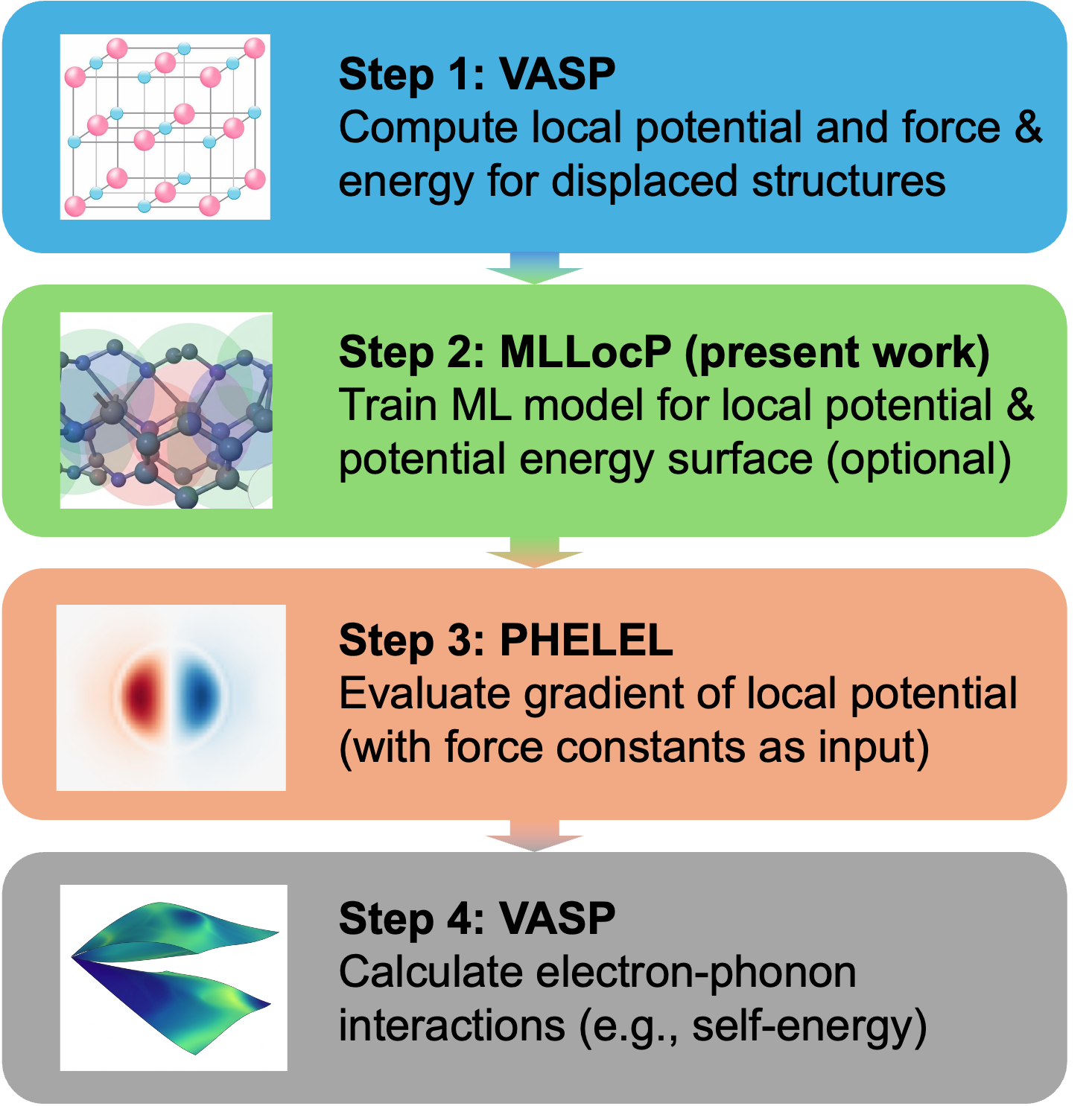}
  \caption{Workflow for machine learning accelerated electron-phonon calculations. 
  }
  \label{fig:workflow}
\end{figure}

\section{Computational workflow}
\label{sec:workflow}

Figure~\ref{fig:workflow} summarizes the proposed workflow. The procedure consists of four stages.

First, reference DFT calculations are performed using VASP for a set of thermally displaced atomic configurations generated either from random displacements or sampled from molecular dynamics trajectories. For each configuration, the smooth local potential, overlap matrices, PAW strength matrices, and total energies and forces are stored for potential ML tasks. In this work, we also compute the displaced configurations generated by PHELEL to provide direct finite-difference derivatives from DFT for validation.

Second, the ML model is trained to map atomic configurations to the local potential on real-space grid points. Instead of using all grid points, targeted and feature-diversity sampling is applied to select representative points. The model is validated on both the potential field itself and its displacement derivatives, because derivative accuracy is the quantity most relevant to electron-phonon coupling.

Third, the trained model is used to generate local potentials for atomic configurations/displacements generated by PHELEL. Displacement derivatives $\partial\tv_{\ml}/\partial R_{\kappa\alpha}$ are then constructed with the PHELEL finite-displacement approach~\cite{Engel2022,PHELEL} [Eq.~\eqref{eq:ml_fd}].

Fourth, the computed derivative potentials are passed to the PAW electron-phonon matrix-element evaluation. The local potential contribution $g^{(V)}$ uses the PHELEL-derived ML derivatives, while terms involving derivatives of the PAW projectors and the remaining augmentation contributions are computed internally within VASP. Phonon frequencies and eigenvectors are obtained from the conventional finite-displacement force-constant approach in PHELEL rather than from an ML interatomic potential; in principle, that phonon step could be replaced by a separately trained ML force field, as now a standard procedure using ML interatomic potentials~\cite{batatia2022mace,batzner20223}, but we do not pursue that route here. Transport quantities such as electron self-energies, relaxation times, and conductivity can then be computed using the same post-processing steps implemented within VASP as in the direct finite-displacement approach.

\section{Computational details}
\label{sec:computational_details}

\subsection{Test systems}

We validate the approach first on elemental fcc Cu and cubic BAs as simple model systems that test the effectiveness of MLLocP for a metal and a polar semiconductor, respectively. We then turn to a 32-atom special quasirandom structure (SQS)~\cite{Zunger1990} of chemically disordered Cu$_3$Au as a low-symmetry disordered alloy, where the reduction in the number of explicit DFT finite-displacement calculations afforded by the ML approach is most consequential. For each system we assess the learned local potential, its displacement derivatives, and the resulting electron-phonon observables against direct DFT finite-displacement references. These cells are not intended as fully converged transport benchmarks, but rather as controlled tests of accuracy and computational savings across chemically and electronically distinct materials.

\subsection{DFT parameters}

All DFT calculations are performed within the PAW method using VASP~\cite{KresseHafner1993,Kresse1996,KresseJoubert1999}. The electron-phonon finite-displacement workflow follows the PHELEL/VASP implementation~\cite{Chaput2019,Engel2022,PHELEL}: both the displacement derivatives of the local potential and the phonon force constants are obtained from finite-displacement supercell calculations~\cite{PHELEL}. We do not train an ML interatomic potential for phonons in this work; phonon frequencies and eigenvectors are taken from the PHELEL/Phonopy~\cite{Togo2015} finite-displacement approach. In principle, that phonon step could be avoided by training a separate ML force field on the same configurations, which is a straightforward extension. The thermalized structures with atomic displacements used for ML training are generated using the quantum covariance matrix constructed using the harmonic force constants~\cite{xia2018revisiting}.

We use the Perdew-Burke-Ernzerhof (PBE) version~\cite{Perdew1996} of the generalized gradient approximation (GGA)~\cite{perdew1996generalized} for the exchange-correlation functional and employ Cu ($3d^{10}4s^{1}$), Au ($5d^{10}6s^{1}$), B ($2s^{2}2p^{1}$), and As ($4s^{2}4p^{3}$) as valence electrons. Plane-wave cutoffs of 520~eV (Cu), 400~eV (Cu$_3$Au), and 420~eV (BAs) are used. The Brillouin zone of each supercell is sampled with KSPACING parameter of 0.2, and electronic occupations use Gaussian smearing of width $\sigma=0.01$~eV for the metals (Cu and Cu$_3$Au) and the tetrahedron method for BAs. The self-consistent local potential is written on a real-space grid of $108^3$ (Cu), $140^3$ (BAs), and $100^3$ (Cu$_3$Au), and the same grid convention is used for every configuration. Finite-displacement derivatives use a central-difference amplitude of 0.03~\AA. The electron-phonon matrix elements, self-energies, and transport quantities are evaluated with the PHELEL/VASP workflow using an electronic $k/q$-mesh of 80$^3$ (Cu and BAs) and 10$^3$ (Cu$_3$Au).

We do not include spin-orbit coupling (SOC) in the present calculations. SOC has been shown to be important for hole transport in BAs~\cite{Liu2018BAs} and in Si~\cite{Ma2018Si}, where it lifts valence band degeneracies and can change phonon-limited hole mobilities substantially. In addition, the electron-phonon transport wave-vector sampling for the disordered Cu$_3$Au SQS is limited to a $10^3$ $k/q$ mesh by the memory bottleneck of the VASP electron-phonon evaluation for this 32-atom cell using our available computational resources. Because SOC is omitted and the alloy transport mesh is comparatively coarse, we do not expect quantitative agreement with experiment for the absolute transport coefficients reported here. The same DFT settings are used, however, for both the direct finite-displacement reference and the MLLocP workflow, so that the comparisons of transport properties below remain a controlled validation of the learned local potential rather than a prediction of experimental transport.

\subsection{ML model and training protocol}
\label{sec:ml_details}

The local potential model follows the mapping in Eq.~\eqref{eq:ml_mapping} and adopts the FIREANN architecture~\cite{Feng2025}, adapted here to predict the self-consistent local potential rather than the electron density. Each real-space grid point is treated as a virtual atom and mapped to a scalar output through the EAD features of Eq.~\eqref{eq:ead_feature}. The same FIREANN hyperparameters are used for Cu, BAs, and Cu$_3$Au. 
The atomic environment is built from eight Gaussian-type orbitals up to maximum angular momentum $L_{\max}=2$ within a cutoff radius $r_c=6$\,$r_{\rm Bohr}$ ($r_{\rm Bohr}$=0.529~\AA)~for Cu and Cu$_3$Au and $r_c=8$\,$r_{\rm Bohr}$~for BAs. These correspond to approximately 3.17 and 4.23~\AA{}, respectively, and are smaller than half of the corresponding supercell dimensions (7.26~\AA{} for Cu, 9.60~\AA{} for BAs, and 7.57~\AA{} for Cu\(_3\)Au). Periodic boundary conditions are treated by explicitly generating lattice translations and including all periodic atomic images lying within \(r_c\); thus, the local environments do not contain repeated self-images of the type that would arise from an overly large cutoff.
The orbital-coefficient network that generates the embedding coefficients $c_\kappa$ [Eq.~\eqref{eq:ead_feature}] is used with $T=0$ message-passing iterations (a single embedding pass without recursive updates). Two feed-forward sub-networks are employed: one predicting the grid-point output with two hidden layers of width $128$ and one predicting the orbital coefficients with two hidden layers of width $64$.

Within the FIREANN framework, we use the efficient grid-point sampling strategy introduced by Feng \textit{et al.}~\cite{Feng2025}. The method combines targeted sampling with linearly independent feature-based sampling to select training points from the dense PAW potential grid. The model is trained to predict the local potential $\tv^{(s)}(\rr_g)$, retaining $N_{\mathrm{sam}}=1000$ points per configuration out of $N_g=108^3$ (Cu), $140^3$ (BAs), and $100^3$ (Cu$_3$Au), corresponding to sampling fractions of approximately $0.08\%$, $0.04\%$, and $0.1\%$, respectively. Following Ref.~\cite{Feng2025}, optimization uses the AdamW algorithm~\cite{AdamW} with batch size $1$ and an initial learning rate of $10^{-3}$ that is decayed by a factor of $0.5$ whenever the validation error does not decrease for $50$ epochs; training is stopped once the learning rate drops below $10^{-6}$, with a maximum of $5000$ epochs.

For each system the training and test partitions are drawn at the level of atomic configurations to avoid leakage between nearby grid points of the same structure (Table~\ref{tab:dataset}). Throughout this work we adopt a fixed $90\%/10\%$ train/test split of the sampled configurations: for a total of $N_s$ thermally displaced structures, $0.9N_s$ are used for training and $0.1N_s$ for testing (e.g., $N_s=10$ gives $9$ training and $1$ test structures). Cu and BAs each use $N_s=10$ configurations ($9/1$), while the chemically disordered Cu$_3$Au SQS uses $N_s=40$ ($36/4$), reflecting the larger configurational diversity of the alloy cell. All accuracy metrics and figures in Sec.~\ref{sec:results} are evaluated on the held-out test partition for each system. 
To assess the statistical robustness of the model, we additionally performed four independent training calculations for each system using different random train/test partitions and neural-network initialization seeds. Full three-dimensional-grid prediction errors are highly consistent among these independent realizations; the detailed statistics are reported in the Supplemental Material (Sec. I). The representative configurations and planes shown below are used primarily to visualize the spatial structure of the prediction errors.
We emphasize that the quantity of ultimate interest is the displacement derivative $\partial \tv/\partial R_{\kappa\alpha}$; we therefore report both field-level and derivative-level accuracy in Sec.~\ref{sec:results}.

\begin{table}[t]
\caption{Dataset partition for each test system. Configurations are split $90\%/10\%$ into disjoint training and test sets at the level of atomic structures; all reported accuracies refer to the held-out test partition.}
\label{tab:dataset}
\begin{ruledtabular}
\begin{tabular}{l c c c c}
System & Cell (atoms) & Grid & $N_s$ & Train/Test \\
\colrule
Cu & 32 & $108^3$ & 10 & 9/1 \\
BAs & 64 & $140^3$ & 10 & 9/1 \\
Cu$_3$Au (SQS) & 32 & $100^3$ & 40 & 36/4 \\
\end{tabular}
\end{ruledtabular}
\end{table}

\begin{figure*}[htp]
    \includegraphics[width = 0.85\linewidth]{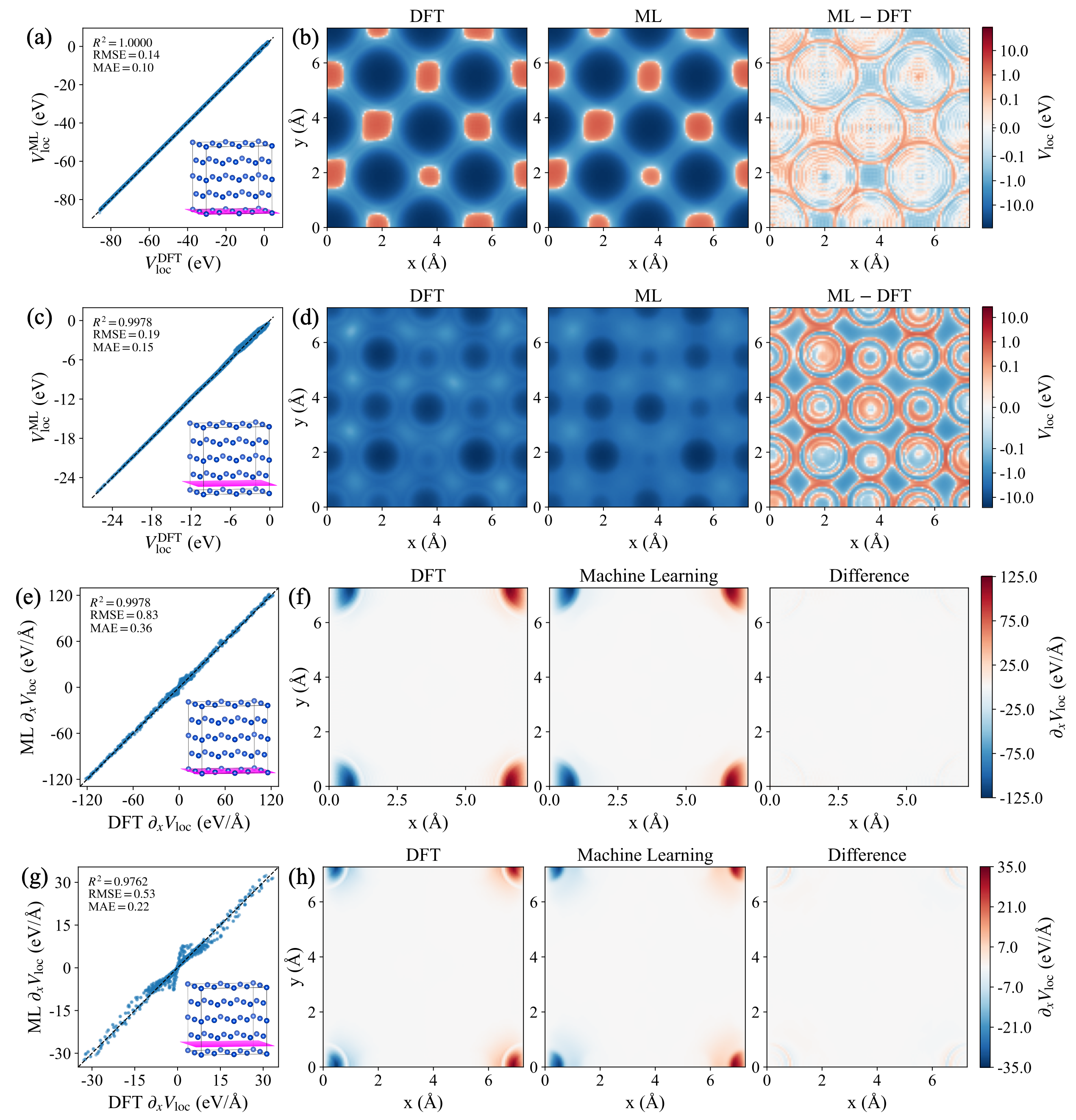}
    \caption{
    Comparison of the ML and DFT local potentials, $V_{\mathrm{loc}}$, and their spatial derivatives, $\partial_x V_{\mathrm{loc}}$, for a thermally displaced 32-atom conventional Cu supercell. (a) Parity plot comparing the ML prediction, $V_{\mathrm{loc}}^{\mathrm{ML}}$, with the DFT reference, $V_{\mathrm{loc}}^{\mathrm{DFT}}$, evaluated at all real-space grid points on the selected $xy$ plane at the fractional coordinate $z=0$ within a $108\times108\times108$ grid, as illustrated by the crystal structure and plane shown in the inset. The dashed line indicates perfect agreement, while the coefficient of determination, $R^2$, root-mean-square error (RMSE), and mean absolute error (MAE) are reported in the inset. (b) Corresponding real-space contour maps of the DFT and ML local potentials and their pointwise difference, $V_{\mathrm{loc}}^{\mathrm{ML}}-V_{\mathrm{loc}}^{\mathrm{DFT}}$. The potential magnitude is represented using a symmetric logarithmic color scale.
    (c) and (d) Same as (a) and (b), respectively, but evaluated on a different selected $xy$ plane at the fractional coordinate $z=0.125$. (e) and (f) Corresponding derivatives of the local potentials with respect to the displacement of a Cu atom along the $x$ axis shown in (a) and (b), respectively.  (g) and (h) Corresponding derivatives of the local potentials shown in (c) and (d), respectively.
    }
    \label{fig:cu_locv}
\end{figure*}

\begin{figure*}[htp]
    \includegraphics[width = 0.75\linewidth]{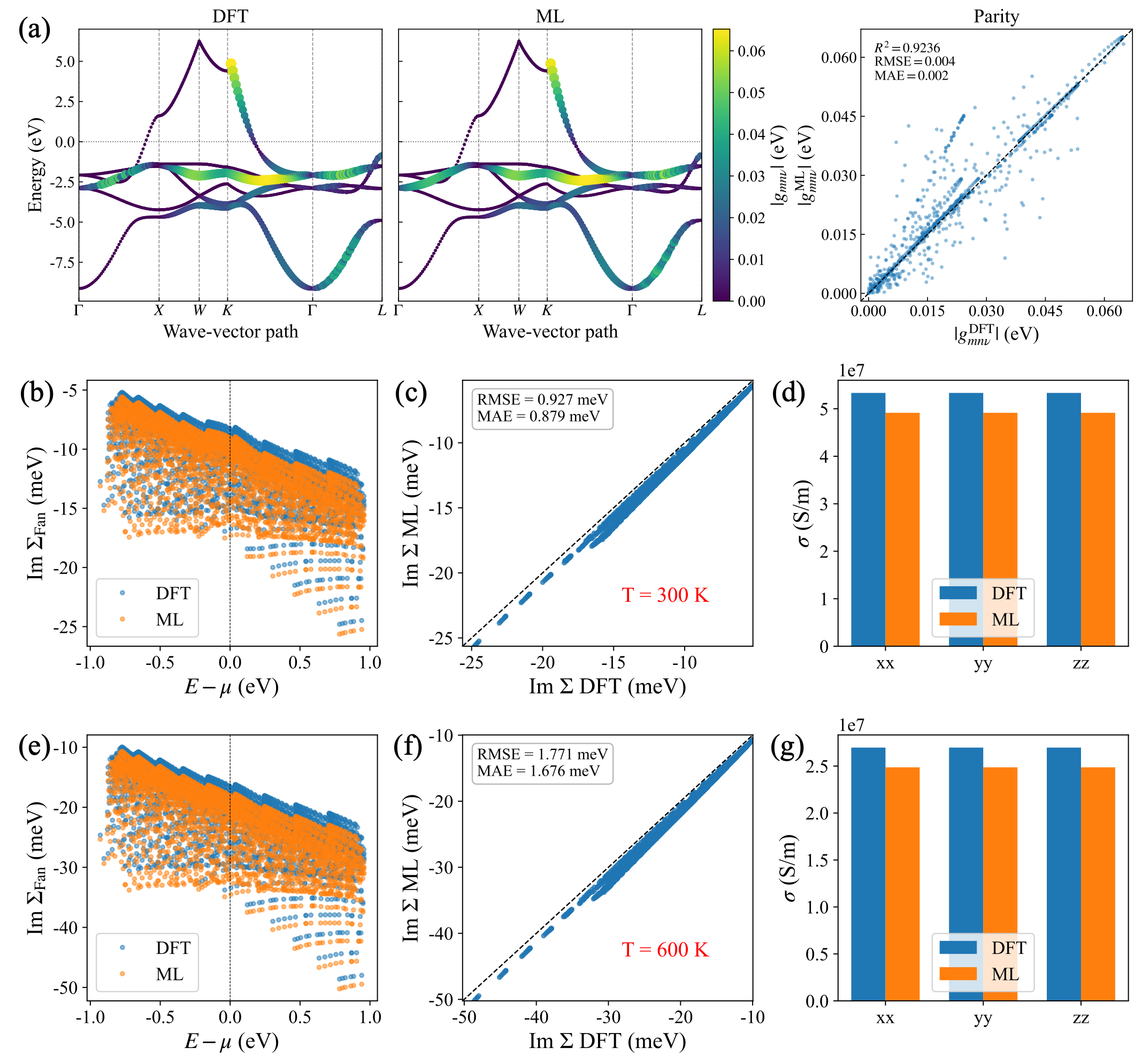}
    \caption{
    (a) Electron-phonon matrix elements along a high-symmetry
    $k{+}q$ path for Cu.
    (Left, center)
    Electronic band structures relative to the Fermi energy,
    with marker color and size indicating the magnitude of the
    electron-phonon coupling $|g_{mn\nu}|$ for a selected
    initial band (the 6th band from the lowest band shown), initial $k$ point ($\Gamma$ point), and phonon mode $\nu$ (the lowest acoustic mode),
    comparing DFT (left) and ML (center) predictions on a
    common color scale.
    Thin gray lines show the underlying bands; vertical dashed
    lines mark high-symmetry points and the horizontal dotted
    line denotes the Fermi level.
    (Right)
    Parity plot of $|g_{mn\nu}^{\mathrm{ML}}|$ versus
    $|g_{mn\nu}^{\mathrm{DFT}}|$ for the same selection,
    with the dashed line indicating perfect agreement and
    inset statistics ($R^{2}$, RMSE, and MAE).
    (b)
    Comparison of the imaginary part of the Fan self-energy,
    $\mathrm{Im}\,\Sigma_{\mathrm{Fan}}$, as a function of
    energy relative to the Fermi energy $\mu$ for Cu from DFT and ML calculations within the self-energy relaxation-time approximation (SERTA) at $T=300\,\mathrm{K}$.
    (c)
    Parity plot of $\mathrm{Im}\,\Sigma$ from ML versus DFT,
    with the dashed line indicating perfect agreement and
    reported RMSE and MAE.
    (d)
    Diagonal components of the electrical conductivity tensor
    $\sigma_{xx}$, $\sigma_{yy}$, and $\sigma_{zz}$ obtained
    from the DFT and ML self-energies.
    (e), (f), and (g) Same as (b), (c) and (d), respectively, but at $T=600\,\mathrm{K}$.
    }
    \label{fig:cu_trans}
\end{figure*}

\section{Results and discussion}
\label{sec:results}

\subsection{Elemental copper (Cu)}
\label{sec:results_cu}

Figure~\ref{fig:cu_locv}(a-d) compares the ML and DFT local potentials for the held-out test configuration of a 32-atom conventional Cu supercell on a $108^3$ real-space grid. Fig.~\ref{fig:cu_locv}(b) shows $V_\loc^\DFT$ on an $xy$ plane at $z=0$ that coincides with a Cu~(001) atomic plane, where $V_\loc$ spans approximately $-85.9$ to $2.1$~eV. The corresponding parity plot in Fig.~\ref{fig:cu_locv}(a) demonstrates very good agreement between $V_\loc^\ml$ and $V_\loc^\DFT$, with $R^2=1.000$, RMSE $=0.14$~eV, and MAE $=0.10$~eV. This agreement is corroborated by the contour comparison in Fig.~\ref{fig:cu_locv}(b): the pointwise difference $V_\loc^\ml-V_\loc^\DFT$ remains small and exhibits a weak ring-like pattern around the atomic sites, consistent with the smooth, screened character of the metallic local potential away from the PAW augmentation spheres.

Moving the plane into the interstitial region (an $xy$ plane at $z=0.125$) between consecutive Cu layers [inset of Fig.~\ref{fig:cu_locv}(c)] reduces the range of $V_\loc$ to approximately $-26.3$ to $-0.5$~eV. On this plane the parity statistics degrade slightly to $R^2=0.9978$, RMSE $=0.19$~eV, and MAE $=0.15$~eV [Fig.~\ref{fig:cu_locv}(c)]. Relative deviations become more noticeable where $|V_\loc|$ is small, and the contour comparison in Fig.~\ref{fig:cu_locv}(d) accordingly shows emerging differences between the DFT and ML maps. These results indicate that the model is less accurate in the weaker interstitial potential than near the Cu atoms.


Because the electron-phonon matrix element depends on $\partial V_{\rm loc}/\partial R_{\kappa\alpha}$ rather than on $V_{\rm loc}$ itself, derivative accuracy is the decisive test for Cu. Panels (e)--(h) of Fig.~\ref{fig:cu_locv} compare the DFT and ML spatial derivatives $\partial_x V_\loc$ on the same two planes used above. On the Cu~(001) plane at $z=0$, $\partial_x V_\loc$ spans roughly $\pm 125$~eV/\AA{} and is sharply peaked at the atomic sites [Fig.~\ref{fig:cu_locv}(f)]. The corresponding parity plot yields $R^2=0.9978$, RMSE $=0.83$~eV/\AA, and MAE $=0.36$~eV/\AA{} [Fig.~\ref{fig:cu_locv}(e)], and the pointwise difference map is nearly featureless. On the interstitial plane at $z=0.125$ the derivative amplitude is substantially weaker (approximately $\pm 35$~eV/\AA), and the parity statistics degrade to $R^2=0.9762$, RMSE $=0.53$~eV/\AA, and MAE $=0.22$~eV/\AA{} [Fig.~\ref{fig:cu_locv}(g)], with faint residual structure appearing near the atomic sites in the difference map [Fig.~\ref{fig:cu_locv}(h)]. Absolute errors are therefore smaller in the interstitial region, but relative agreement is reduced where the derivative itself is weak, mirroring the trend already seen for $V_\loc$. It is worth noting that there are relatively strong deviations between DFT and ML when $\partial_x V_\loc$ is vanishing [Fig.~\ref{fig:cu_locv}(g)], which could lead to overestimated absolute values of the electron-phonon matrix elements. Overall, the learned derivatives remain faithful to the DFT fields at the finite-difference amplitude $\delta=0.03$~\AA{} used in Eq.~\eqref{eq:ml_fd}, indicating that our MLLocP strategy is reasonably accurate for subsequent PAW sandwich evaluations.


Figure~\ref{fig:cu_trans}(a) examines how these derivative errors propagate into $|g_{mn\nu}|$. We consider a high-symmetry $\kk{+}\qq$ path for a selected initial electron state (the sixth band from the lowest band shown at $\Gamma$ point) and the lowest acoustic phonon branch. The DFT and ML band structures, color-coded by coupling strength on a common scale from 0.0 to 0.06~eV, are visually indistinguishable across the high-symmetry path. The parity plot of $|g_{mn\nu}^\ml|$ versus $|g_{mn\nu}^\DFT|$ for the same selection yields $R^2=0.9236$, RMSE $=0.004$~eV, and MAE $=0.002$~eV. Although with a relatively high $R^2$, the parity plot shows appreciable scattering about the diagonal. We find that the magnitude of this scattering behavior varies with the selected electronic bands and phonon branches.  This trend is consistent with our earlier observation that larger relative errors tend to occur when $V_{\rm loc}$ is weak.
We note that these values of $|g_{mn\nu}|$ are considerably smaller than those in BAs, indicating weaker electron-phonon interactions in Cu.
This trend is consistent with our earlier observation that larger relative errors tend to occur when $V_{\rm loc}$ is weak.


Figure~\ref{fig:cu_trans}(b-g) reports calculated transport quantities obtained from the ML and DFT electron-phonon couplings within the self-energy relaxation-time approximation (SERTA)~\cite{Li2015,Gunst2016,PHELEL}. At $T=300$~K, the imaginary part of the Fan-Migdal self-energy $\mathrm{Im}\,\Sigma_\mathrm{Fan}(E-\mu)$ from ML closely tracks the DFT reference over the window $|E-\mu|\lesssim 1$~eV wherein $\mu$ denotes the Fermi level of Cu [Fig.~\ref{fig:cu_trans}(b)], and the corresponding parity plot gives RMSE $=0.927$~meV and MAE $=0.879$~meV [Fig.~\ref{fig:cu_trans}(c)]. The ML self-energy is slightly and systematically larger in its absolute magnitude than DFT, which carries through to the conductivity: the diagonal components $\sigma_{xx}$, $\sigma_{yy}$, and $\sigma_{zz}$ are isotropic, with DFT values of approximately $5.3\times 10^{7}$~S/m and ML values of approximately $4.9\times 10^{7}$~S/m, an underestimation of about $8\%$ [Fig.~\ref{fig:cu_trans}(d)]. At $T=600$~K the same pattern persists with larger absolute scattering rates [Fig.~\ref{fig:cu_trans}(e)]: the parity errors rise to RMSE $=1.771$~meV and MAE $=1.676$~meV [Fig.~\ref{fig:cu_trans}(f)], while the conductivity drops to approximately $2.7\times 10^{7}$~S/m (DFT) and $2.5\times 10^{7}$~S/m (ML), again within about $7\%$ [Fig.~\ref{fig:cu_trans}(g)]. Therefore, MLLocP preserves not only the local field and its derivatives but also the integrated metallic transport property of Cu, with relative conductivity errors below 10\%. 
Additional tests show that this systematic bias is reproducible across independent training realizations and persists when the finite-displacement amplitude is varied from 0.01 to 0.06~\AA{}. The dominant correlation is instead with residual errors in the calculated local-potential derivative where the DFT derivative is small: the ML model tends to produce a small nonzero \(|\partial V_{\rm loc}/\partial R|\) [Fig.~3(g)], which slightly enhances the resulting electron-phonon scattering rates and lowers the conductivity. Detailed convergence tests are provided in the Supplemental Material (Sec. II).

\subsection{Boron arsenide (BAs)}
\label{sec:results_bas}

\begin{figure*}[htp]
    \includegraphics[width = 0.85\linewidth]{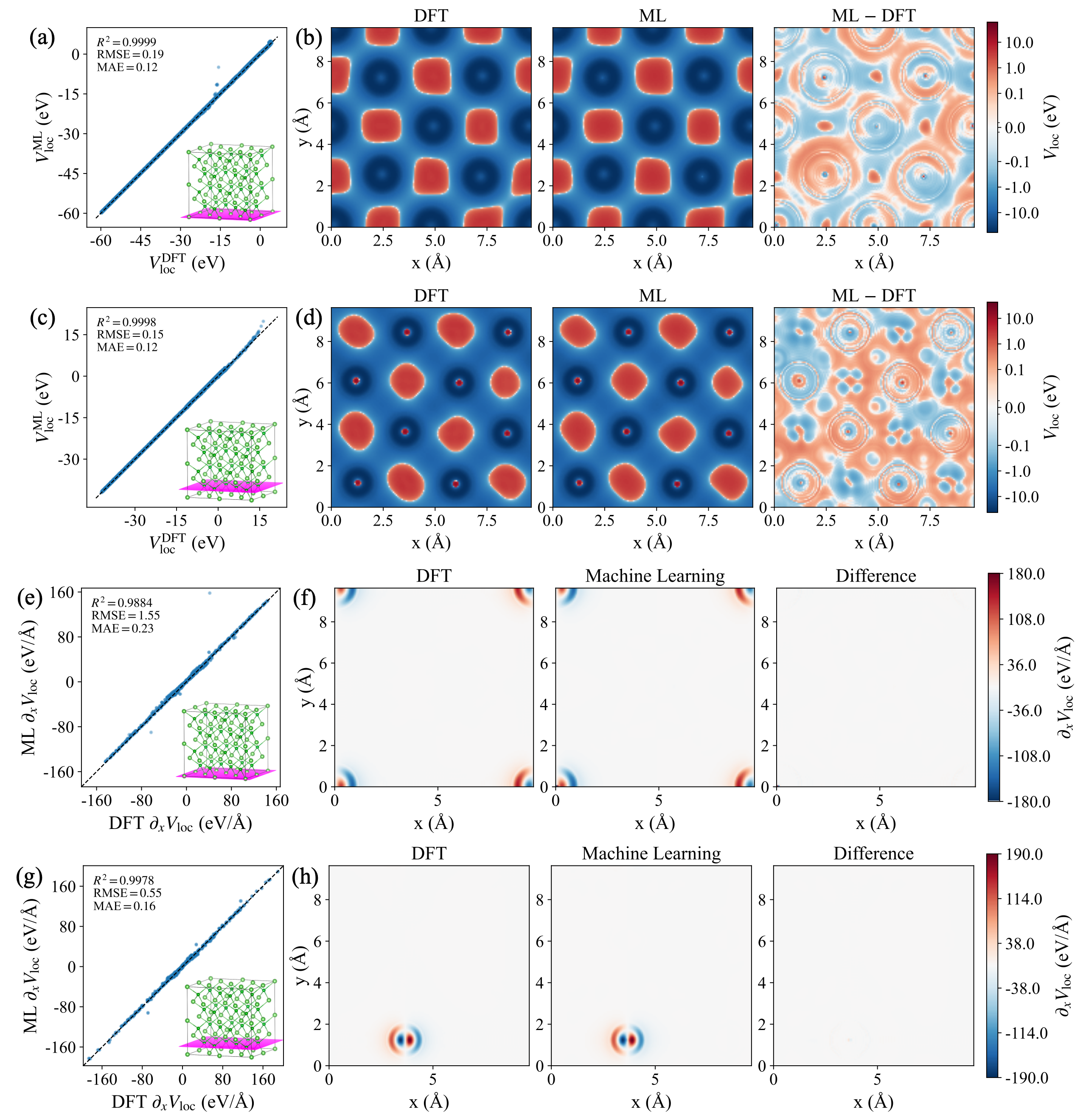}
    \caption{
    Comparison of the ML and DFT local potentials, $V_{\mathrm{loc}}$, and their spatial derivatives, $\partial_x V_{\mathrm{loc}}$, for a thermally displaced 64-atom conventional BAs supercell. (a) Parity plot comparing the ML prediction, $V_{\mathrm{loc}}^{\mathrm{ML}}$, with the DFT reference, $V_{\mathrm{loc}}^{\mathrm{DFT}}$, evaluated at all real-space grid points on the selected $xy$ plane at the fractional coordinate $z=0$ within a $140\times140\times140$ grid, as illustrated by the crystal structure and plane shown in the inset. The dashed line indicates perfect agreement, while $R^2$, RMSE, and MAE are reported in the inset. (b) Corresponding real-space contour maps of the DFT and ML local potentials and their pointwise difference, $V_{\mathrm{loc}}^{\mathrm{ML}}-V_{\mathrm{loc}}^{\mathrm{DFT}}$. The potential magnitude is represented using a symmetric logarithmic color scale.
    (c) and (d) Same as (a) and (b), respectively, but evaluated on a different selected $xy$ plane at the fractional coordinate $z=0.125$. (e) and (f) Corresponding derivatives of the local potentials with respect to the displacement of a As atom along the $x$ axis shown in (a) and (b), respectively.  (g) and (h) Corresponding derivatives of the local potentials with respect to the displacement of a B atom along the $x$ axis shown in (c) and (d), respectively.
    }
    \label{fig:bas_locv}
\end{figure*}

We next turn to cubic BAs as a polar semiconductor. Despite the polar character of BAs, we train directly on the full self-consistent local potential without subtracting an analytic long-range dipole contribution. Fig.~\ref{fig:bas_locv}(a-d) compares the ML and DFT $V_\loc$ on a $140^3$ real-space grid. On the $xy$ plane at $z=0$, which intersects an atomic layer of As atoms, $V_\loc$ spans a broad range of roughly $-60$ to $5$~eV [Fig.~\ref{fig:bas_locv}(a)]. The parity plot yields $R^2=0.9999$, RMSE $=0.19$~eV, and MAE $=0.12$~eV [Fig.~\ref{fig:bas_locv}(a)], and the DFT and ML contour maps are visually indistinguishable, including the deep potential wells at the atomic sites. Shifting the plane to $z=0.125$ where B atoms reside changes the sampled environments and the dynamic range of $V_\loc$ (approximately $-45$ to $25$~eV), yet the agreement remains good: $R^2=0.9998$, RMSE $=0.15$~eV, and MAE $=0.12$~eV [Fig.~\ref{fig:bas_locv}(c,d)]. However, there are increased discrepancies between ML and DFT towards the large positive values of $V_\loc$. Relative to Cu, the field-level errors are comparable in absolute terms even though BAs presents sharper covalent features and a polar charge distribution. These results show that a purely local FIREANN model can already reproduce the self-consistent potential of this polar cell without an explicit short-range/long-range decomposition. This is due to the usage of a larger cutoff distance (8~$r_{\rm Bohr}$) among atoms and probes that can cover the 64-atom cell dimension.


Panels (e)-(h) of Fig.~\ref{fig:bas_locv} compare $\partial_x V_\loc$. On the $z=0$ plane the derivative field exhibits large, tightly localized bipolar lobes near As sites, with amplitudes of order $\pm 160$~eV/\AA{} [Fig.~\ref{fig:bas_locv}(e)]. The parity statistics are $R^2=0.9884$, RMSE $=1.55$~eV/\AA, and MAE $=0.23$~eV/\AA{} [Fig.~\ref{fig:bas_locv}(e)]; the comparatively large RMSE probably arises from the scattered points with large deviations. On the second plane, which crosses the B atoms the bipolar features remain sharp and the overall agreement improves to $R^2=0.9978$, RMSE $=0.55$~eV/\AA, and MAE $=0.16$~eV/\AA{} [Fig.~\ref{fig:bas_locv}(g)], and the difference maps are essentially featureless [Fig.~\ref{fig:bas_locv}(g)]. The cutoff in the ML model is therefore already sufficient to capture the displacement response of the local potential in this 64-atom polar cell.

\begin{figure*}[htp]
    \includegraphics[width = 0.75\linewidth]{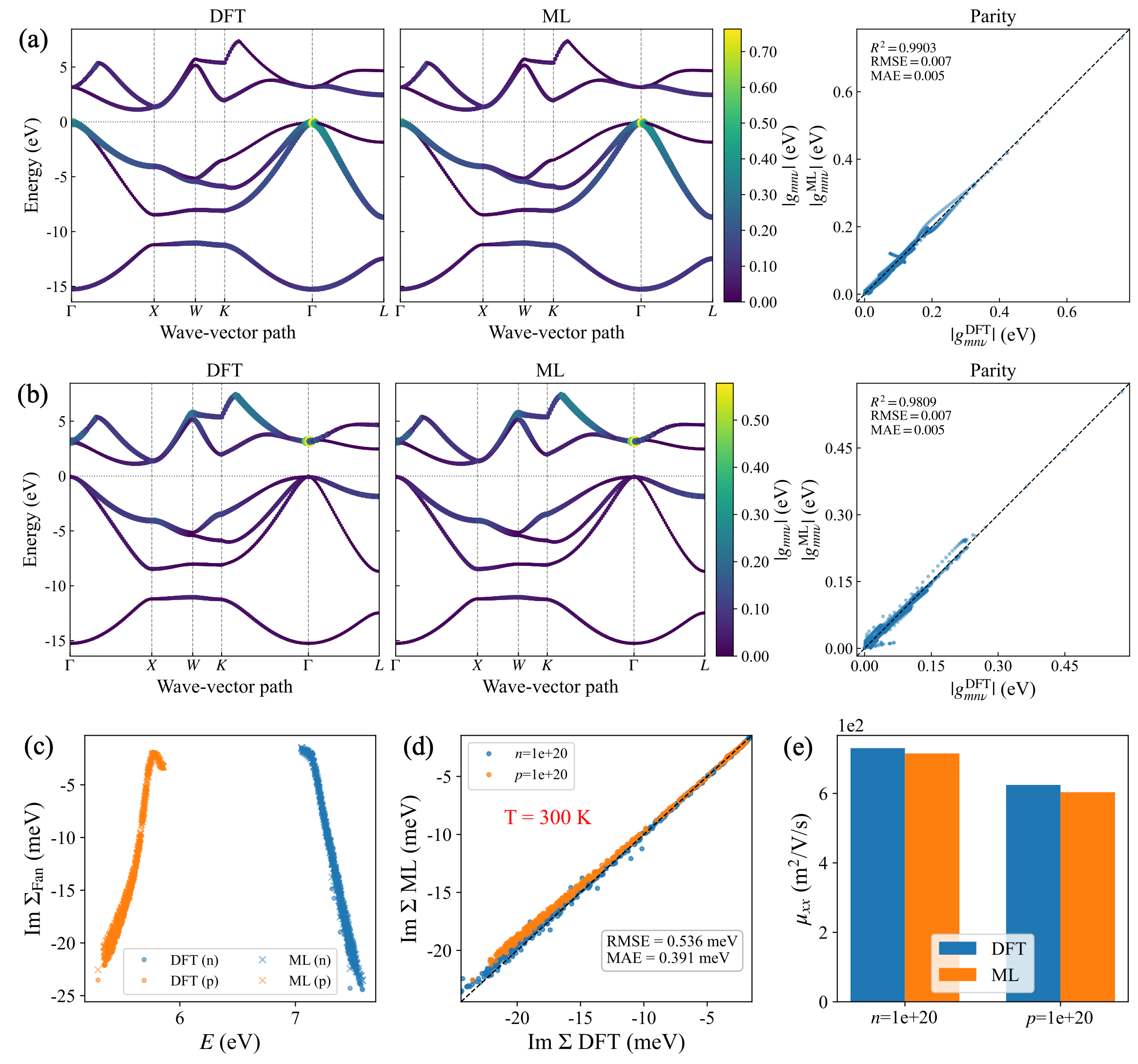}
    \caption{
    (a) Electron-phonon matrix elements along a high-symmetry
    $k{+}q$ path for BAs.
    (Left, center)
    Electronic band structures relative to the valence band maximum (VBM),
    with marker color and size indicating the magnitude of the
    electron-phonon coupling $|g_{mn\nu}|$ for a selected
    initial band (VBM), initial $k$ point ($\Gamma$ point), and phonon mode $\nu$ (longitudinal optical mode),
    comparing DFT (left) and ML (center) predictions on a
    common color scale.
    Thin gray lines show the underlying bands; vertical dashed
    lines mark high-symmetry points and the horizontal dotted
    line denotes the Fermi level.
    (Right)
    Parity plot of $|g_{mn\nu}^{\mathrm{ML}}|$ versus
    $|g_{mn\nu}^{\mathrm{DFT}}|$ for the same selection,
    with the dashed line indicating perfect agreement and
    inset statistics ($R^{2}$, RMSE, and MAE).
    (b) Same as (a) but for the conduction band minimum (CBM).
    (c)
    Comparison of the imaginary part of the Fan self-energy,
    $\mathrm{Im}\,\Sigma_{\mathrm{Fan}}$, as a function of
    energy for both holes and electrons for BAs from DFT and ML calculations within the self-energy relaxation-time approximation (SERTA) at $T=300\,\mathrm{K}$.
    (d)
    Parity plot of $\mathrm{Im}\,\Sigma$ from ML versus DFT,
    with the dashed line indicating perfect agreement and
    reported RMSE and MAE.
    (e)
    Diagonal components of the electron and hole mobility tensor
    $\mu_{xx}$ obtained
    from the DFT and ML self-energies at electron and hole concentrations of $10^{20}~{\rm cm}^{-3}$.
    }
    \label{fig:bas_trans}
\end{figure*}

Figure~\ref{fig:bas_trans}(a) examines $|g_{mn\nu}|$ along a high-symmetry $\kk{+}\qq$ path for the valence band maximum (VBM) at $\Gamma$ coupled to the longitudinal optical (LO) mode, a channel that is particularly sensitive to polar scattering. The DFT and ML coupling patterns on the band structure are in close visual agreement, and the parity plot gives $R^2=0.9903$, RMSE $=0.007$~eV, and MAE $=0.005$~eV. Fig.~\ref{fig:bas_trans}(b) repeats the analysis for the conduction band minimum (CBM), yielding $R^2=0.9809$, RMSE $=0.007$~eV, and MAE $=0.005$~eV. Couplings in BAs are substantially larger  (up to about $0.7$~eV for the VBM-LO selection) than the acoustic mode matrix elements of Cu, yet the learned local potential derivatives remain accurate for both hole- and electron-related states.


Figure~\ref{fig:bas_trans}(c-e) reports hole and electron transport at $T=300$~K within SERTA. The energy-resolved $\mathrm{Im}\,\Sigma_\mathrm{Fan}$ for $n$- and $p$-type carriers at a concentration of $10^{20}\,\mathrm{cm}^{-3}$ shows near-perfect overlap between DFT and ML [Fig.~\ref{fig:bas_trans}(c)], and the corresponding parity plot yields RMSE $=0.536$~meV and MAE $=0.391$~meV [Fig.~\ref{fig:bas_trans}(d)], which are smaller than those found for Cu despite the polar character of BAs. The diagonal mobility $\mu_{xx}$ obtained from the ML self-energies reproduces the DFT values for both electrons and holes to within 3.5\% percent [Fig.~\ref{fig:bas_trans}(e)]. Together with the field-level results above, this confirms that MLLocP captures the polar electron-phonon physics of BAs through to carrier mobility without a separate long-range correction for the present cell size.

\subsection{Chemically disordered Cu$_3$Au}
\label{sec:results_cu3au}

Finally, we consider a 32-atom special quasirandom structure of chemically disordered Cu$_3$Au. This cell breaks translational symmetry and presents inequivalent Cu and Au local environments, so it is the most stringent field-level test among the three systems and the primary target for computational savings from MLLocP.

\begin{figure*}[htp]
    \includegraphics[width = 0.85\linewidth]{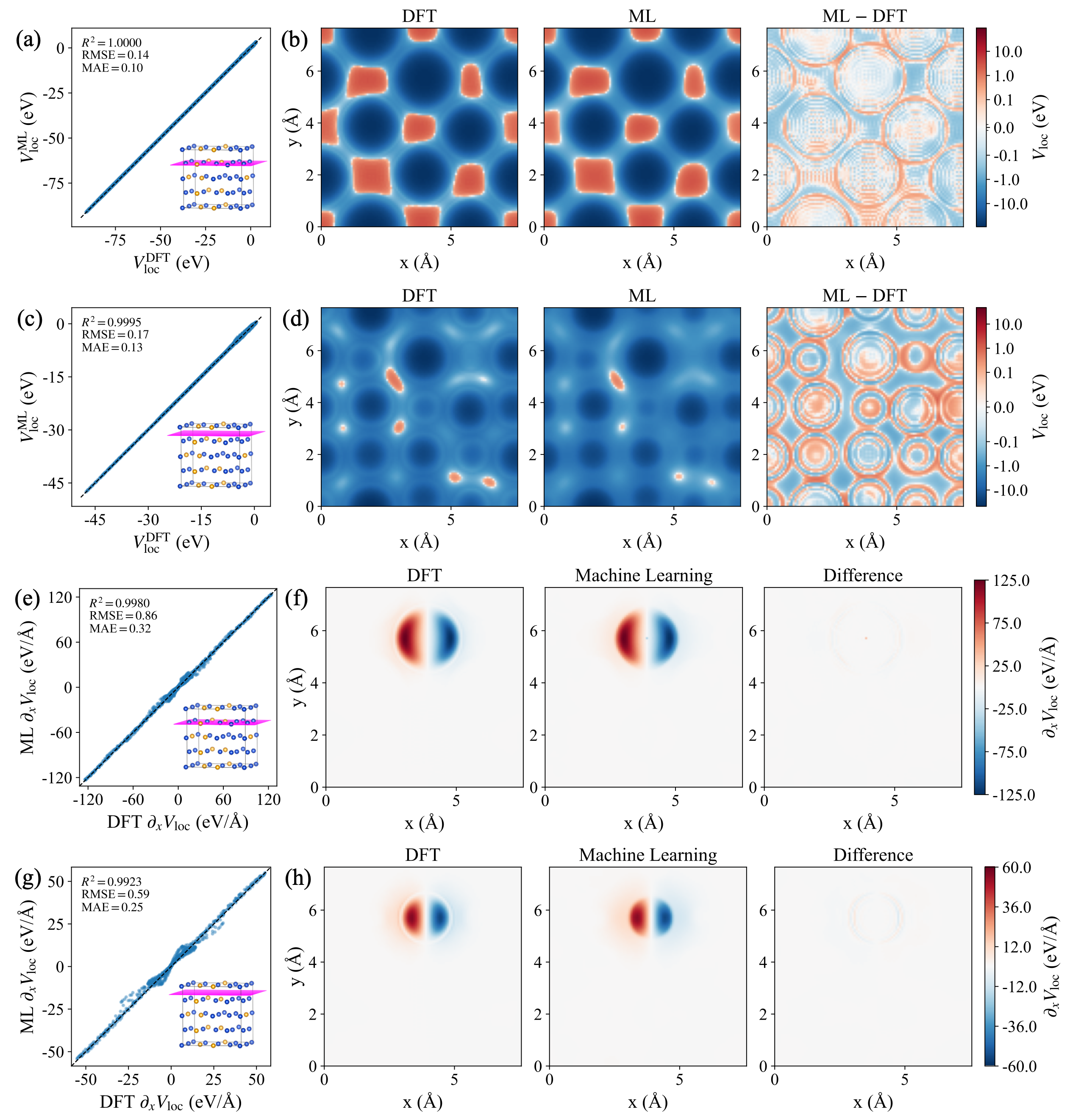}
    \caption{
    Comparison of the ML and DFT local potentials, $V_{\mathrm{loc}}$, and their spatial derivatives, $\partial_x V_{\mathrm{loc}}$, for a thermally displaced 32-atom disordered Cu$_3$Au structure. (a) Parity plot comparing the ML prediction, $V_{\mathrm{loc}}^{\mathrm{ML}}$, with the DFT reference, $V_{\mathrm{loc}}^{\mathrm{DFT}}$, evaluated at all real-space grid points on the selected $xy$ plane at the fractional coordinate $z=0.743$ within a $100\times100\times100$ grid, as illustrated by the crystal structure and plane shown in the inset. The dashed line indicates perfect agreement, while $R^2$, RMSE, and MAE are reported in the inset. (b) Corresponding real-space contour maps of the DFT and ML local potentials and their pointwise difference, $V_{\mathrm{loc}}^{\mathrm{ML}}-V_{\mathrm{loc}}^{\mathrm{DFT}}$. The potential magnitude is represented using a symmetric logarithmic color scale.
    (c) and (d) Same as (a) and (b), respectively, but evaluated on a different selected $xy$ plane at the fractional coordinate $z=0.868$. (e) and (f) Corresponding derivatives of the local potentials with respect to the displacement of a Au atom along the $x$ axis  shown in (a) and (b), respectively.  (g) and (h) Corresponding derivatives of the local potentials shown in (c) and (d), respectively.
    }
    \label{fig:cu3au_locv}
\end{figure*}

Figure~\ref{fig:cu3au_locv}(a-d) compares the ML and DFT local potentials on a $100^3$ grid. On the $xy$ plane at fractional coordinate $z=0.743$, which intersects a mixed Cu/Au layer, $V_\loc$ spans roughly $-90$ to $5$~eV and exhibits species-dependent well depths at the atomic sites [Fig.~\ref{fig:cu3au_locv}(b)]. The parity plot gives $R^2=1.0000$, RMSE $=0.14$~eV, and MAE $=0.10$~eV [Fig.~\ref{fig:cu3au_locv}(a)], matching the field-level accuracy obtained for elemental Cu on the atomic plane [Fig.~\ref{fig:cu_locv}(a)]. On a second plane at $z=0.868$, which lies between two neighboring atomic planes, the range is reduced (approximately $-50$ to $0$~eV) and the statistics remain high: $R^2=0.9995$, RMSE $=0.17$~eV, and MAE $=0.13$~eV [Fig.~\ref{fig:cu3au_locv}(c,d)]. Absolute field errors are therefore comparable to those for Cu, indicating that the $N_s=40$ sample set ($36$ train / $4$ test) captures alloy-specific spatial structure, including environments that differ from pure Cu because of local Au neighbors, from the same FIREANN training protocol.


Panels (e)-(h) of Fig.~\ref{fig:cu3au_locv} compare $\partial_x V_\loc$. On the $z=0.743$ plane the derivative field shows strong bipolar features near atomic sites, with amplitudes of order $\pm 125$~eV/\AA{} [Fig.~\ref{fig:cu3au_locv}(f)]. The parity statistics are $R^2=0.9980$, RMSE $=0.86$~eV/\AA, and MAE $=0.32$~eV/\AA{} [Fig.~\ref{fig:cu3au_locv}(e)], essentially matching the corresponding Cu~(001) derivative accuracy. On the $z=0.868$ plane the derivative amplitude decreases (approximately $\pm 60$~eV/\AA) and the agreement is $R^2=0.9923$, RMSE $=0.59$~eV/\AA, and MAE $=0.25$~eV/\AA{} [Fig.~\ref{fig:cu3au_locv}(g,h)], again with only faint residuals in the difference maps. Because the SQS breaks translational symmetry, Cu- and Au-centered displacements sample distinct chemical environments; the present results show that a single model learns both species-dependent responses. A direct central-difference DFT calculation for this 32-atom cell would require $192$ self-consistent runs [Eq.~\eqref{eq:disp_count}], so the demonstrated derivative fidelity is the key prerequisite for the cost reduction discussed in Sec.~\ref{sec:cost}.

\begin{figure*}[htp]
    \includegraphics[width = 0.75\linewidth]{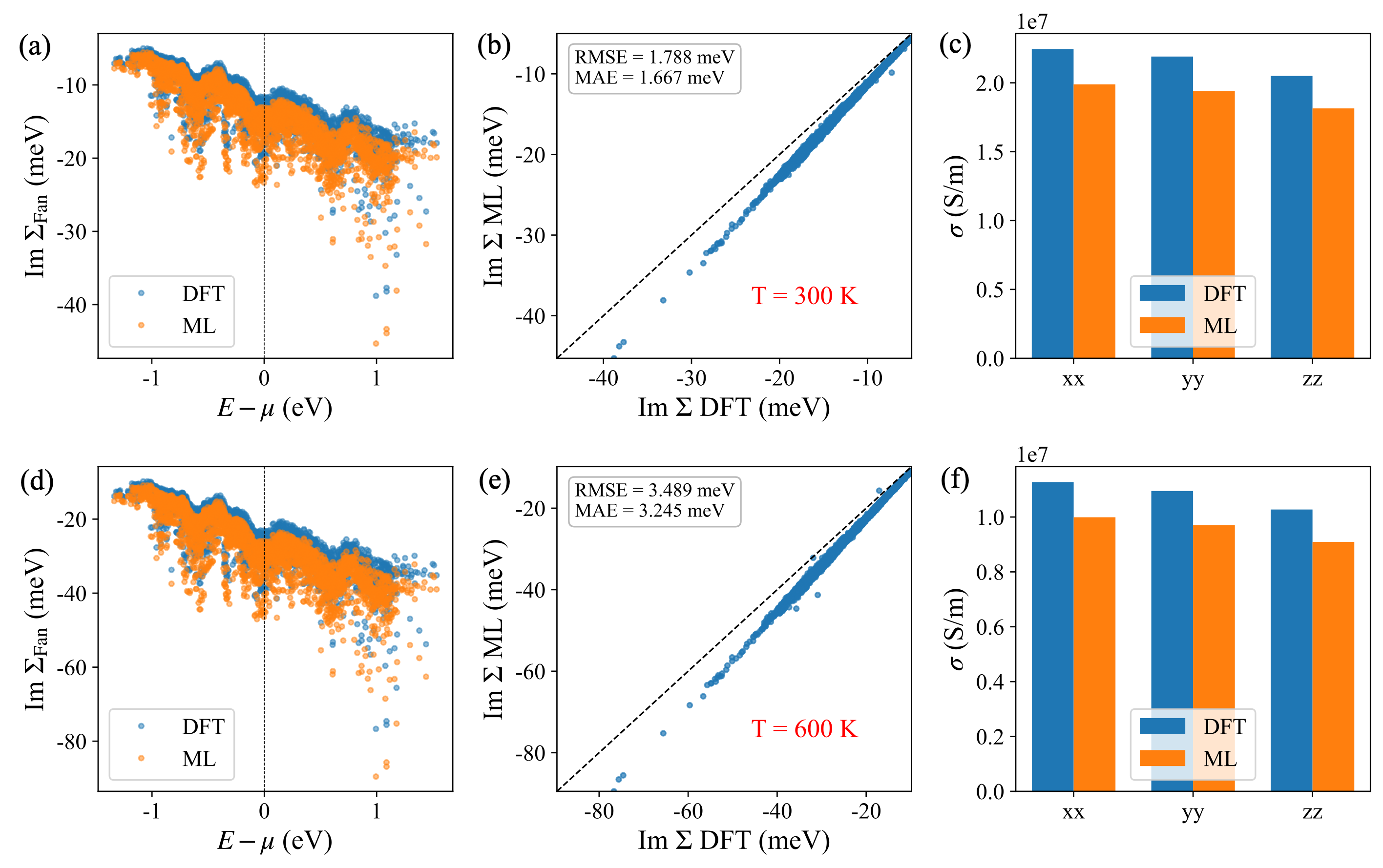}
    \caption{
    (a)
    Comparison of the imaginary part of the Fan self-energy,
    $\mathrm{Im}\,\Sigma_{\mathrm{Fan}}$, as a function of
    energy relative to the chemical potential $\mu$ for disordered Cu$_3$Au from DFT and ML calculations within the self-energy relaxation-time approximation (SERTA) at $T=300\,\mathrm{K}$.
    (b)
    Parity plot of $\mathrm{Im}\,\Sigma$ from ML versus DFT,
    with the dashed line indicating perfect agreement and
    reported RMSE and MAE.
    (c)
    Diagonal components of the electrical conductivity tensor
    $\sigma_{xx}$, $\sigma_{yy}$, and $\sigma_{zz}$ obtained
    from the DFT and ML self-energies.
    (d)--(f) Same as (a)--(c), respectively, but at $T=600\,\mathrm{K}$.
    }
    \label{fig:cu3au_trans}
\end{figure*}

Figure~\ref{fig:cu3au_trans} reports SERTA transport for the same SQS Cu$3$Au. Alloy transport in a finite SQS cell is not intended to represent the thermodynamic disorder limit; the comparison here tests whether ML-derived local-potential derivatives preserve the end-to-end carrier transport properties of this specific disordered cell. At $T=300$~K, $\mathrm{Im}\,\Sigma_\mathrm{Fan}(E-\mu)$ from ML tracks the DFT distribution over $|E-\mu|\lesssim 1.5$~eV [Fig.~\ref{fig:cu3au_trans}(a)], with parity RMSE $=1.788$~meV and MAE $=1.667$~meV [Fig.~\ref{fig:cu3au_trans}(b)]. Relative to elemental Cu the absolute self-energy errors remain somewhat larger, and the ML values lie systematically slightly below the DFT parity line (more negative $\mathrm{Im}\,\Sigma$), corresponding to mildly enhanced scattering. This bias carries through to the conductivity: the diagonal components $\sigma_{xx}$, $\sigma_{yy}$, and $\sigma_{zz}$ retain the weak SQS anisotropy of the DFT reference, and ML underestimates each component by roughly $11\%$ [Fig.~\ref{fig:cu3au_trans}(c)], e.g., $\sigma_{xx}\approx 2.25\times 10^{7}$~S/m (DFT) versus $\approx 2.0\times 10^{7}$~S/m (ML) at 300~K.

At $T=600$~K the scattering rates roughly double [Fig.~\ref{fig:cu3au_trans}(d)], and the parity errors scale accordingly to RMSE $=3.489$~meV and MAE $=3.245$~meV [Fig.~\ref{fig:cu3au_trans}(e)]. The same systematic overestimation of $|\mathrm{Im}\,\Sigma|$ persists, and the conductivity remains lower than DFT by a similar relative margin of about $11$--$12\%$ across $\sigma_{xx}$, $\sigma_{yy}$, and $\sigma_{zz}$ [Fig.~\ref{fig:cu3au_trans}(f)]. Thus, using $N_s=40$ configurations, MLLocP reproduces the energy- and temperature-dependent metallic transport properties of this chemically disordered alloy with a conductivity error of approximately $11\%$. This accuracy is comparable to the $\sim 8\%$ bias observed for elemental Cu, while avoiding most of the $192$ self-consistent finite-difference calculations.

To test whether the present framework depends on the relatively regular environments of the cubic systems considered above, we additionally applied MLLocP to a monoclinic carbon allotrope with space group \(P2/c\). The learned full-grid local potential achieves \(R^2=0.9997\) and RMSE \(=0.316\) eV, while the resulting Fan self-energy and anisotropic conductivity components remain in good agreement with direct DFT calculations. The complete results are presented in the Supplemental Material (Sec. III).

\subsection{Derivatives of PAW strength}
\label{sec:gD_vs_gV}

The central approximation of this work replaces the local potential term $g^{(V)}$ by a learned surrogate while retaining $g^{(P)}$ and $g^{(R)}$ from the PAW formalism. For all three test systems we find that the PAW strength contribution $g^{(D)}$ is negligible compared with $g^{(V)}$. This finding is consistent with the PAW finite-displacement analysis of Chaput \textit{et al.}~\cite{Chaput2019}, who reported that $g^{(D)}$ and $g^{(R)}$ are negligibly small for aluminium. 
The small magnitude of $g^{(D)}$ indicates that the PAW strength parameters respond only weakly to small atomic displacements. For Cu in our study, for example, our numerical analysis shows that their magnitudes change by less than $0.25\%$. This weak variation is physically reasonable because these parameters primarily describe the near-core electronic structure and the all-electron reconstruction within the PAW augmentation regions. We therefore set $g^{(D)}$ to zero in electron-phonon calculations that use ML local potentials, while retaining it in the corresponding DFT references. 
As a direct test, setting \(g^{(D)}=0\) also in the DFT reference changes the calculated Cu conductivity by only approximately 1.1\%, further confirming its negligible impacts.
The end-to-end agreement for $|g_{mn\nu}|$, $\mathrm{Im}\,\Sigma$, and transport in Cu, BAs, and Cu$_3$Au further supports this hierarchy: once $g^{(V)}$ is accurately learned, the absence of derivatives of the PAW strength does not limit the final observables. We therefore do not develop a separate ML model for the PAW strength matrix in this work, leaving such an extension to future studies of systems in which $g^{(D)}$ may play a more significant role.

\begin{figure}[t]
    \includegraphics[width = 0.95\linewidth]{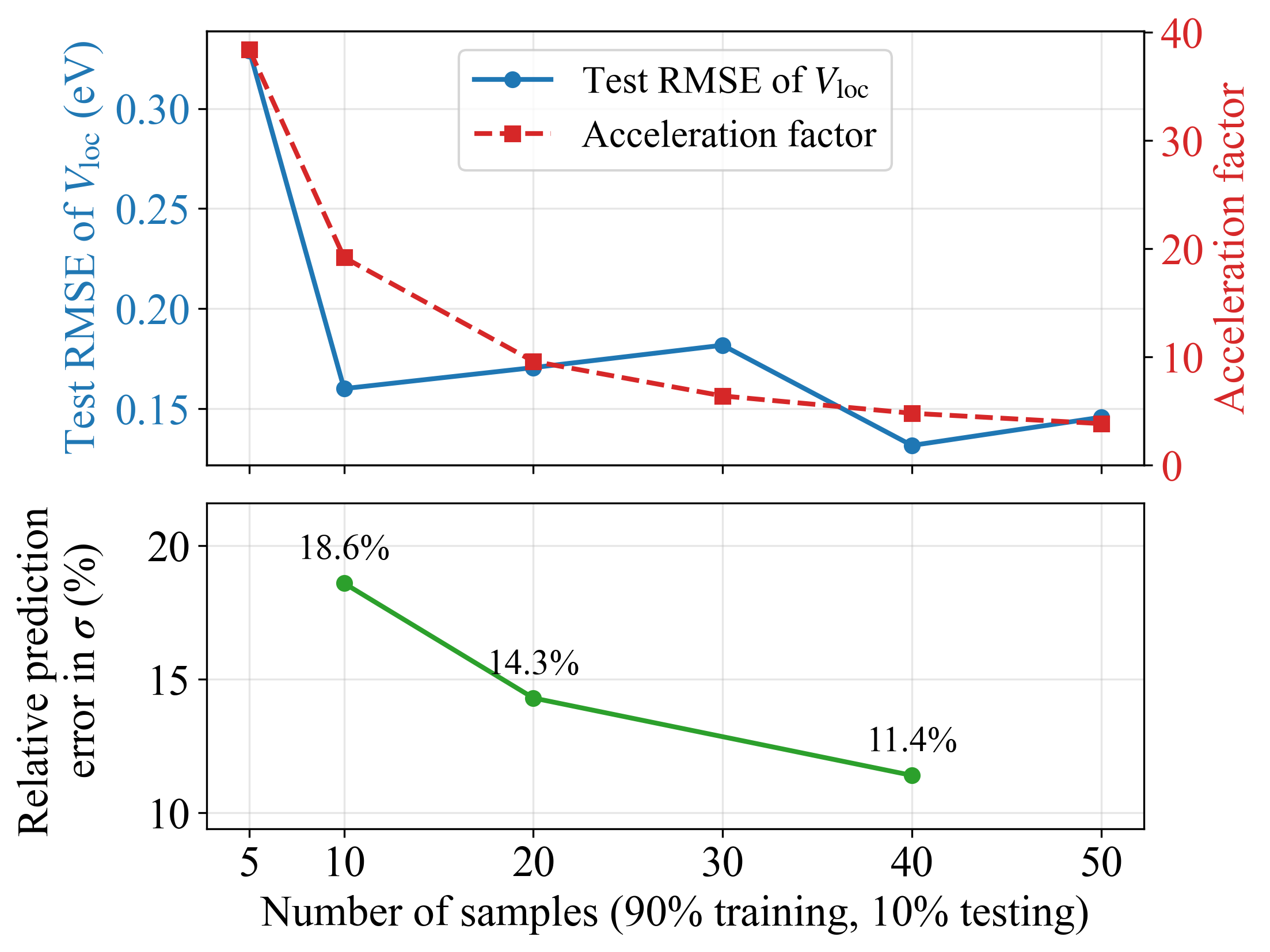}
    \caption{
    Accuracy-cost trade-off for chemically disordered Cu$_3$Au as a function of the number of sampled configurations $N_s$, always partitioned $90\%/10\%$ into training and testing (e.g., $N_s=10$ gives $9$ train / $1$ test). (a)~Test-set RMSE of $V_{\mathrm{loc}}$ (left axis, blue) and acceleration factor $S=N_{\mathrm{disp}}/N_s$ relative to a full central-difference calculation with $N_{\mathrm{disp}}=192$ (right axis, red). (b)~Relative prediction error in the electrical conductivity $\sigma$ with respect to DFT, annotated for $N_s=10$ ($18.6\%$), $20$ ($14.3\%$), and $40$ ($11.4\%$).
    }
    \label{fig:cu3au_speedup}
\end{figure}

\subsection{Computational cost and speedup}
\label{sec:cost}

For high-symmetry crystals such as fcc Cu and cubic BAs, crystal symmetry substantially reduces the number of independent atomic displacements. A conventional finite-difference electron-phonon calculation therefore does not require the full set of $6N$ supercell calculations. The main computational challenge instead arises in low-symmetry and chemically disordered systems, as illustrated by the case of Cu$3$Au. Without symmetry reduction, a central-difference calculation displaces each of the $N$ atoms along $\pm x$, $\pm y$, and $\pm z$, leading to $N_{\mathrm{disp}} = 6N = 192$ self-consistent DFT calculations for the 32-atom alloy cell. Each displaced configuration requires a complete self-consistent evaluation of the local potential. As a result, the $6N$ scaling dominates the computational cost whenever symmetry cannot be exploited.

Our MLLocP framework avoids most of these self-consistent calculations by inferring the local potential from a model trained on a subset of $N_s$ reference configurations. Fig.~\ref{fig:cu3au_speedup} summarizes the resulting trade-off between accuracy and computational cost for Cu$_3$Au, including both the prediction error in local potential  and the relative error in the predicted electrical conductivity. We define the acceleration factor as the reduction in the number of self-consistent DFT calculations needed to generate the sampled configurations:
\begin{equation}
S = \frac{N_{\rm disp}}{N_s} = \frac{192}{N_s}.
\label{eq:disp_count}
\end{equation}
This definition gives $S\approx 19$, $10$, and $5$ for $N_s=10$, $20$, and $40$, respectively (Fig.~\ref{fig:cu3au_speedup}, top panel, right axis). Over the same range, the test RMSE of $V_\loc$ decreases from approximately $0.33$~eV at $N_s=5$ to approximately $0.16$~eV at $N_s=10$. It then remains within a relatively narrow range of approximately $0.13$-$0.18$~eV and decreases relatively slowly for $N_s=20$-$50$ (Fig.~\ref{fig:cu3au_speedup}, top panel, left axis). The field-level learning curve therefore begins to saturate near ten sampled configurations, whereas the acceleration factor continues to decrease as $1/N_s$.

The relative prediction error in conductivity converges gradually with increasing sample set size (Fig.~\ref{fig:cu3au_speedup}, bottom panel). Relative to DFT, the prediction error in conductivity is $18.6\%$ for $N_s=10$, $14.3\%$ for $N_s=20$, and $11.4\%$ for the production choice of $N_s=40$. Thus, $N_s=40$ provides an acceleration factor of $S\approx 5$ while maintaining a conductivity error of approximately $11\%$, comparable to that obtained for elemental Cu. Smaller sample sets offer greater computational savings. In particular, reducing $N_s$ to $20$ or $10$ increases the acceleration to approximately 10-fold or 20-fold, respectively, while raising the relative conductivity error to $14.3\%$ or $18.6\%$. These intermediate choices remain useful when greater acceleration is preferred and a modest reduction in transport accuracy is acceptable.

Eq.~\eqref{eq:disp_count} counts only the self-consistent DFT calculations required to generate the $N_s$ labeled configurations. It does not include the cost of ML training and inference, both of which are comparatively small once the reference potentials have been generated. Because the number of sampled configurations is expected to grow much more slowly than $6N$ once the relevant local environments are adequately represented, the projected acceleration should increase for larger disordered and defect-rich supercells. These are precisely the systems for which direct finite-displacement electron-phonon calculations are currently most computationally demanding.

\section{Conclusion and outlook}
\label{sec:conclusion}

We have developed and demonstrated MLLocP, a strategy for accelerating PAW-based finite-displacement electron-phonon calculations by using machine learning to predict the self-consistent local potential on real-space grids. Starting from the PAW decomposition of the electron-phonon matrix element into local potential, PAW strength, projector derivative, and reconstruction contributions, we identified the local potential derivative $\partial\tv/\partial R_{\kappa\alpha}$ as the most natural first target. This quantity is high-dimensional and determined self-consistently, yet it remains a smooth scalar field governed by the same locality principles that underpin modern charge-density learning methods. We therefore adapt the FIREANN architecture and its efficient grid-point sampling strategy~\cite{Feng2025} from electron-density learning to local potential learning.

Across a simple metal (Cu), a polar semiconductor (BAs), and a chemically disordered alloy (Cu\(_3\)Au), the learned potentials and displacement derivatives closely reproduce their DFT counterparts. Independent train/test partitions and random initializations yield consistent full-grid errors, confirming that this accuracy is not specific to a particular data split. Additional validation for a monoclinic carbon allotrope further demonstrates that the approach is not restricted to cubic structures.
We also find that the PAW strength contribution $g^{(D)}$ is negligible, consistent with previous PAW finite-displacement analyses~\cite{Chaput2019}. After propagation through the standard PAW sandwich operations, the learned derivatives reproduce electron-phonon matrix elements with RMSE values of only a few meV and Fan self-energies with parity errors ranging from sub-meV to a few meV. The resulting transport properties also remain accurate. MLLocP reproduces the electrical conductivities of Cu and Cu$_3$Au to within approximately $8\%$ and $11\%$, respectively, and the carrier mobility of BAs to within 3.5\%. For Cu$_3$Au, the method reduces the number of self-consistent finite-difference calculations from $192$ to an $N_s=40$ sample set, consisting of $36$ training and $4$ test configurations under a $90\%/10\%$ split. In the present work, phonon frequencies and eigenvectors are obtained from PHELEL finite-displacement calculations rather than from a machine learning force field.

It is also useful to place MLLocP in the context of recent Hamiltonian-learning approaches~\cite{DeepH2022,HamGNN2023,zhong2024accelerating} such as DeepH~\cite{DeepH2022} and its extension to deep-learning density functional perturbation theory~\cite{li2024first}. These methods learn the electronic Hamiltonian directly in a localized atomic-orbital representation and can therefore replace a substantially larger fraction of the underlying first-principles calculation, offering the potential for greater computational acceleration. MLLocP instead adopts a complementary and more conservative strategy: only the self-consistent local potential is learned on the native PAW real-space grid, while the pseudo wave functions, PAW projectors, overlap matrices, and all-electron reconstruction terms remain within the established VASP/PHELEL formalism. The present speedup is consequently more moderate, but the approach avoids introducing an additional localized electronic basis and preserves the native PAW representation in the final electron-phonon matrix elements. These two directions are therefore complementary, with Hamiltonian-learning methods emphasizing maximal acceleration and MLLocP emphasizing direct integration with plane-wave PAW electron-phonon calculations and minimal modification of the underlying first-principles framework. We also note that, most recently, the DeepH framework has been reformulated to directly learn the the real-space Kohn-Sham potential (named DeepH-R~\cite{yuan2026deep}), sharing a similar spirit as our MLLocP framework.

The framework also points to several promising directions for future development. First, FIREANN is not the only architecture suitable for MLLocP. Other machine learning models, including state-of-the-art equivariant graph neural networks~\cite{batzner20223,batatia2022mace}, could address the same local potential learning problem. Equivariant representations are especially well suited to rotationally covariant quantities. They could therefore serve not only as alternatives to FIREANN for learning the scalar local potential, but also as direct models for the PAW strength matrices $D_{ij}$ and their displacement derivatives. These tensorial one-center quantities carry higher-order rotational structure and are currently retained from DFT rather than learned.

Second, combining MLLocP with a separately trained machine learning interatomic potential would eliminate the remaining first-principles finite-displacement calculations needed to obtain phonon frequencies, eigenvectors, and force constants. Such a combination would enable a fully machine learning accelerated electron-phonon workflow. Third, extending the learned local potential to spin-polarized and spin-orbit coupled systems would generalize $\tv$ from a scalar field to a multicomponent quantity, opening the method to magnetic, topological, and heavy-element materials. Fourth, strongly polar, low-dimensional, and heterogeneous systems will likely require an explicit treatment of the long-range response mediated by Born effective charges. In such cases, the electrostatic perturbation extends beyond the finite real-space cutoff used in the present model~\cite{Verdi2015,Sjakste2015}.

Another important goal is to construct transferable models that learn local potential responses across multiple compositions, structures, and bonding environments rather than within a single material class. Tighter integration of training and inference with VASP and PHELEL would also remove the remaining file-transfer and grid-processing overhead and allow predicted local-potential derivatives to enter electron-phonon calculations directly. Together, these advances could transform MLLocP from a material specific acceleration strategy into a broadly transferable platform for first-principles electron-phonon calculations in large, low-symmetry, defective, alloyed, and disordered systems that remain prohibitively expensive for conventional finite-displacement approaches.

\section*{Declaration of generative AI and AI-assisted technologies in the manuscript preparation process}

During the preparation of this work, the author(s) used ChatGPT in order to polish language and improve readability. After using this tool/service, the author(s) reviewed and edited the content as needed and take(s) full responsibility for the content of the published article.

\begin{acknowledgments}
\textbf{Acknowledgments:} 
Y. X. acknowledges 1) the support from the U.S. National Science Foundation through award No.~DMR-2532261 and 2) the computing resources provided by Bridges2 at Pittsburgh Supercomputing Center (PSC) through allocations mat220006p and mat220008p from the Advanced Cyber-infrastructure Coordination Ecosystem: Services \& Support (ACCESS) program, which is supported by National Science Foundation grants 2138259, 2138286, 2138307, 2137603, and 2138296. Y. X. is grateful to Z. J. W. and Soda, the cat, for their encouragement and support throughout the preparation of this manuscript. 
\end{acknowledgments}

\bibliography{refs}

\end{document}